\documentclass[aps,nofootinbib,notitlepage,superscriptaddress,twocolumn,10pt,prd]{revtex4-2}

\usepackage{bm}
\usepackage{graphicx}
\usepackage{amsmath,amssymb}
\usepackage{hyperref}
\usepackage{braket}
\usepackage{subfigure}
\usepackage{cleveref}
\usepackage{float}
\usepackage[dvipsnames]{xcolor}
\usepackage{soul}
\usepackage{rotating}
\usepackage{multirow}
\usepackage{mathtools}
\usepackage{lipsum}
\usepackage{makecell}

\usepackage[margin=0.75in]{geometry}

\hypersetup{
	colorlinks=true,
	linkcolor=red,
	citecolor=blue,
}
\usepackage{physics}

\usepackage{natbib}
\usepackage{verbatim}
\usepackage{dcolumn}
\usepackage[amssymb]{SIunits}
\usepackage{tabularx}
\usepackage{booktabs}
\usepackage[normalem]{ulem}

\newcommand{\be}{\begin{equation}}
\newcommand{\ee}{\end{equation}}
\newcommand{\ba}{\begin{eqnarray}}
\newcommand{\ea}{\end{eqnarray}}

\definecolor{ForestGreen}{RGB}{36,179,0}

\newcommand{\avg}[1]{\langle #1 \rangle}
\DeclareMathOperator{\Avg}{Avg}

\usepackage{aas_macros}

\begin{document}
\newcommand\wls{\texttt{WL2}}
\newcommand\all{\texttt{WL23\_WPH\_CMBWL}}

\title{Dimensionality Reduction Techniques for Statistical Inference in Cosmology}

\author{Minsu Park}
\affiliation{Center for Particle Cosmology, Department of Physics and Astronomy, University of Pennsylvania, Philadelphia, PA 19104, USA}
\author{Marco Gatti}
\affiliation{Center for Particle Cosmology, Department of Physics and Astronomy, University of Pennsylvania, Philadelphia, PA 19104, USA}
\author{Bhuvnesh Jain}
\affiliation{Center for Particle Cosmology, Department of Physics and Astronomy, University of Pennsylvania, Philadelphia, PA 19104, USA}
%
\begin{abstract}
We explore linear and non-linear dimensionality reduction techniques for statistical inference of parameters in cosmology. Given the  importance of compressing the increasingly complex data vectors used in cosmology, we address  questions that impact the constraining power achieved, such as: Are currently used methods effectively lossless? Under what conditions do non-linear methods, typically based on neural nets, outperform linear methods? 
Through theoretical analysis and experiments with simulated weak lensing data vectors we compare three standard linear methods and neural network based methods. We also propose two linear methods inspired by information theory: a variation of the MOPED algorithm we call e-MOPED and an adaptation of Canonical Correlation Analysis (CCA), which is a method uncommon in cosmology but well known in statistics. Both e-MOPED and CCA utilize simulations spanning the full parameter space, and rely on the sensitivity of the data vector to the parameters of interest. In the context of our experiments, the two linear methods we propose consistently outperform the rest. The gains  are significant compared to compression methods used in the literature: up to 30\% in the Figure  of Merit for $\Omega_m$ and $S_8$ in a realistic Simulation Based Inference analysis that includes statistical and systematic errors. We also recommend two modifications that improve the performance of all methods: 
First,  include components in the compressed data vector that may not  target the key parameters but still enhance the constraints on due to their correlations. The gain is significant, above 20\% in the Figure of Merit. Second,  compress Gaussian and non-Gaussian statistics separately -- we include two summary statistics of each type in our analysis.  
\end{abstract}

\maketitle

\section{Introduction} \label{Sec:Intro}

Dimensionality reduction (DR), or compression (used interchangeably throughout this paper), is crucial in statistical inference tasks across the sciences. For its use in cosmology, consider the inverse covariance matrix which is necessary when assuming a Gaussian likelihood. Its reliability depends on the length of the data vector and the number of simulations used to estimate the covariance \citep{Hartlap2007,Percival2021}. While it is theoretically possible to obtain a stable covariance matrix for any length of data vector with enough simulations, this is often practically computationally unfeasible.

Simulation-Based Inference (SBI) is increasingly popular \citep{Hahn2017,Levasseur2017,Coogan2020,Wagner-Carena2021,Hermans2021,Wagner-Carena2022,Montel2023,Lemos2023,Massara2024,Abell2024,jeffrey2024dark,gatti_wph1,gatti_wph2, hoffmann2022minimizing}, leveraging machine learning techniques like neural density estimation (NDE) to infer the parameter space posteriors directly from a suite of simulations. While SBI does not explicitly require estimating a covariance matrix (and its inverse), it still relies on mock data to learn the likelihood surface. However, SBI often struggles with long data vectors, as high-dimensional likelihood surfaces are challenging for neural networks to learn reliably without a very large, often unfeasible, number of simulations and mock data.~\citep{yang2022does, hoffmann2022minimizing}

Yet, with the goal of extracting every bit of non-Gaussian as well as Gaussian information from cosmological observables, such as weak lensing maps, the number of widely used summary statistics (and therefore, the length of the data vectors) grows every year \citep{G20,moments2021,Porth2021,Martinet2018,HD2022,Thiele2020,Parroni2020,Parroni2021,Heydenreich2022,Valogiannis2022,Cheng2024,Allys2020,Anbajagane2023,Banerjee2023,ivanov2023cosmology,EuclidNG}. It is crucial, then, to find suitable DR techniques so that statistical inference from the long data vectors of summary statistics can take place in a reliable and computationally efficient way. We focus here on finding DR techniques that perform best under the framework of SBI with N-body simulated what the fk data vectors. However, our results have broad applicability to cases where theory predictions are easier to obtain, or where the likelihood is Gaussian. Methodology developed here will be useful in enhancing the constraints of cosmological parameters with upcoming surveys such as the Vera C. Rubin Observatory Legacy Survey of Space and Time~\citep{ivezic2019lsst}, Euclid~\citep{laureijs2011euclid}, and the Nancy Grace Roman Space Telescope~\citep{spergel2015wide}.

We consider the following linear and non-linear DR techniques. First, we look at two linear techniques that have been widely used for decades: Principal Component Analysis (PCA; \citep{Francis1992, Connolly1995, Madgwick2002, Lahav2009, Zuercher2024}) and Massively Optimized Parameter Estimation (MOPED; \cite{heavens2000massive}). We also consider Canonical Correlation Analysis (CCA), which is well-known but uncommon in cosmology. Additionally, we introduce a new variation of MOPED that significantly reduces the computational cost, which we call e-MOPED.
There are also three neural network based non-linear techniques that have been introduced more recently that differ in the loss functions they employ: in particular, we consider a MSE loss \citep{Jeffrey2021,jeffrey2024dark,gatti_wph1,gatti_wph2}, Variational Mutual Information Maximization \citep{barber2004algorithm}, and Information Maximizing Neural Networks \citep{charnock2018automatic,Makinen2024}. 

There are other compression ideas recently discussed in the astrophysics literature (e.g., \citep{Akhmetzhanova:2023hiy, fewermocks,templatebank, modalcompression, cobraoptimal}). However, some of these approaches are either not general enough to take into account  non-Gaussian statistics, or assume analytical tractability of the data vector and covariance. Therefore, they are not well suited to the problems considered in this work. 
Recently, the authors of \citep{lanzieri2024optimal} investigated optimal neural compression for weak lensing full-field SBI. Our analysis focuses on inference from summary statistics rather than full-field data, and also highlights the power of linear compressors, which are not considered in \citep{lanzieri2024optimal}. Furthermore, the setup used by Lanzieri et al. is more idealistic than what is considered here, as they work with less noisy simulations (LSST Y10 noise level), less realistic simulations (i.e., lognormal mocks), and only vary cosmological parameters, without considering astrophysical or measurement systematics such as intrinsic alignment or redshift biases.


The paper is organised as follows. In~\Cref{Sec:linear_compressors} and~\Cref{Sec:non-linear_compressors} we review the techniques considered in greater detail. In~\Cref{Sec:data} we describe the simulations, summary statistics, and SBI technique used in our experiments.  We  present our findings in~\Cref{sec:results} in the form of four lessons with regards to the various DR techniques. In~\Cref{subsec:ranking} we demonstrate CCA and e-MOEPD typically result in the tightest constraints on cosmological parameters, meaning they are the most information preserving. Secondly, in many of these techniques, there is a clear mathematical connection between the $\alpha^{\rm th}$ cosmological parameter and the $\alpha^{\rm th}$ component of the compressed data vector. In~\Cref{subsec:components} we demonstrate that components of the compressed data vector unrelated to the parameters of interest (in particular ones related to systematics) are still useful in constraining them.  In~\Cref{subsec:pca_global} we point out that while the community often performs PCA with regards to simulations at fiducial cosmology for DR, PCA with regards to simulations at varied cosmology is able to better preserve information from data vectors. Lastly, we argue that it is often beneficial to compress Gaussian statistics and non-Gaussian statistics separately in~\Cref{subsec:separately}. 
We conclude in~\Cref{Sec:conclusion}. We also provide a pedagogical github repository for the methods highlighted here\footnote{\texttt{github.com/98minsu/CosmoCompression}}.

\section{Linear Methods} \label{Sec:linear_compressors}

All the methods in this section are linear DR techniques that work by defining a compression matrix $U$ applied to data vector $t$ such that the compressed data vector $c$ can be defined as follows:
\begin{equation}
\label{eq:definition_compression}
c = Ut.
\end{equation}

Since we are using an SBI framework, we do not have a theoretical model for our data vectors; instead, each data vector \( t \) is obtained through simulations.

We will define the quantities inline the first time they appear in the text, but refer to ~\Cref{tab:variable_def} for a summary of the main variable definitions used throughout this section. Everywhere in this paper, Roman indices iterate over realizations and Greek indices denote parameters (i.e. the $\alpha^{\rm th}$ parameter). All the parameter vectors live in $\mathbb{R}^n$ and all the data vectors live in $\mathbb{R}^m$.


\begin{table}[h!]
\caption{Variables used and their definitions. \label{tab:variable_def}}
\begin{tabular}{ |  >{\centering\arraybackslash}p{0.07\textwidth} |  >{\centering\arraybackslash}p{0.33\textwidth} |  >{\centering\arraybackslash}p{0.05\textwidth} |} 
 \hline
 Variable & Definition & Size   \\ 
 \hline\hline
 $p_i$ & Samples of varied parameters, see~\Cref{subsec:sims}  for the distribution & $n$  \\  \hline
 $t_i$ & Data vector corresponding to $p_i$  & $m$ \\  \hline
 $c_i$ & Compressed data vector & Varies  \\  \hline
 $N_{\rm sim}$ & Number of $(p_i,t_i)$ pairs & -  \\  \hline
  $P, \ T$ & Matrices with $p_i$ and $t_i$ as columns respectively & $(n,m)$ $\times N_{\rm sim}$ \\  \hline
  $p_f^{\ }$ & Fiducial parameters, see~\Cref{subsec:sims}  & $n$ \\  \hline
 $t_{f,i}^{ \ }$ & $i^{\rm th}$ data vector generated at $p_f^{\ }$, $i\in \{ 1 \dots N_f \}$ & $m$\\  \hline
 $p_{f\pm\delta p_\alpha}$ & $p_f^{\ }$ with $\alpha^{\rm th}$ parameter shifted by $\pm \delta{p_\alpha}$ &$n$\\ \hline
 $t_{f\pm\delta p_\alpha, i}$ & $i^{\rm th}$ data vector generated at $p_{f\pm\delta p_\alpha}$, $i\in \{ 1 \dots N_{\pm}\}$ &$m$\\ \hline
 $t_{f;\alpha}$ & Derivative of the data vector at $p_f$ with respect to the $\alpha^{\rm th}$ parameter&$m$\\ \hline 
 $C_f$ & Covariance matrix of $t_{f,i}^{ \ }$ & $m\times m$\\ \hline
 $C_t$ & Covariance matrix of $t_{i}$ & $m\times m$\\ \hline
 $C_p$ & Covariance matrix of $p_{i}$ & $n\times n$\\ \hline
 $C_{tp}$& Cross-covariance matrix of $t_i$ and $p_i$, $C_{pt} = C_{tp}^T$ & $m \times n$\\ \hline
 $C_l$ & $C_p$ projected to data vector space, defined in~\Cref{eq:cl_def} & $m\times m$ \\ \hline
 $A$ & Average Jacobian $\partial t / \partial p$ estimated with $p_i, \ t_i$ defined in~\Cref{eq:ptod_exact} & $m \times n$\\ \hline
 $A_f$ &  Jacobian $\partial t / \partial p$ at $p_f^{\ }$ via finite difference, defined in~\Cref{eq:fdderivative}& $m \times n$\\ \hline
 $\bar{x}$ & Average of quantity or vector $x$ & - \\ \hline
\end{tabular}
\end{table}

\subsection{PCA} \label{subsec:PCA}

There are two approaches to using PCA for DR. The first variant,  which we will call PCA-f for PCA-fiducial, uses the principal components of the covariance matrix of the data vector at fiducial cosmology $C_f$:
\begin{align}
    C_f = \frac{1}{N_f} \sum_i (t_{f,i}^{ \ } - \bar{t}_{f}^{\ })(t_{f,i}^{ \ } - \bar{t}_{f}^{\ })^T. \label{eq:fidcov}
\end{align}
In the above equation, $N_f$ is the total number of noisy data vectors at fiducial cosmology, $t_{f,i}$ is the $i$-th data vector at fiducial cosmology, and $\bar{t}_{f}^{\ }$ is the mean of $t_{f,i}^{ \ }$. This is the approach adopted by some previous weak lensing works \citep{Zuercher2022,Zuercher2024}. The second approach uses the principal components of the covariance matrix of data vectors across varying cosmologies, which we will call PCA:
\begin{align}
    C_t = \frac{1}{N_{\rm sim}}\sum_i (t_{i} - \bar{t})(t_{i} - \bar{t})^T, \label{eq:cov}
\end{align}
where \( N_{\rm sim} \) is the total number of noisy data vectors, \( t_{i} \) is the \( i \)-th noisy data vector at a given set of parameters (cosmological and others) \( p_i \), and \( \bar{t} \) is the mean of \( t_{i} \). 
Let us denote the principal components of the matrix \( C_f \) or \( C_t \) as \( v_i \). They are found by solving the eigenvalue problem
\begin{align}
    C v_i = \lambda_i v_i \label{eq:pca},
\end{align}
where $\lambda_i$ is the eigenvalue of the $i$-th component. This is also equivalent to 
\begin{align}
    \max_{v} \frac{v^T C v}{v^Tv} 
\end{align}
where the solution of the above equation gives the first eigenvector, and subsequent components can be found by solving the same maximization problem with the constraint that the new vectors are orthogonal to the previous ones.
For both the PCA-f and PCA methods, the compression matrix $U$ in \Cref{eq:definition_compression} is a matrix whose rows are the $n_{\rm PCA}$ principal components $v_i$ with the highest variance:
\begin{align}
    U = (v_1 \ v_2 \ \cdots \ v_{n_{\rm PCA}})^T \label{eq:U}.
\end{align}
The number of components $n_{\rm PCA}$ can be as large as the length of the uncompressed data vector, though one would typically aim to keep fewer.  Choosing the number of components to retain in forming the compression matrix $U$ is somewhat arbitrary and requires testing (in \Cref{sec:results} we choose $n_{\rm PCA} = n$, i.e., the total number of parameters varied in our analysis setup).

To understand the difference between the two methods, let us consider the sources of variation captured in \( C_f \) and \( C_t \). In \( C_f \), the variation is entirely due to noise. However, when compressing data vectors, we are primarily interested in retaining the variations related to changes in the parameters (cosmological and others), and the principal components of \( C_f \) might contain little information about those. On the other hand, in \( C_t \), the variations are due to both parameter changes and noise, making it, in principle, better suited to this compression task, as we will see later. This is illustrated by the correlation matrices shown in~\Cref{fig:cov}.


One potential issue with using the principal components of \( C_t \) for compression is related to the set of parameters \( p_i \) that dominate the variations in the data vectors \( t_i \). For instance, if \( p_i \) is sampled to have a small variance in \( \Omega_m \) and \( S_8 \) but a very large variance in the intrinsic alignment parameters \( \eta_{\rm IA} \) and \( A_{\rm IA} \), then \( C_t \) will be dominated by the variations in \( \eta_{\rm IA} \) or \( A_{\rm IA} \) and may contain little to no information on \( \Omega_m \) and \( S_8 \). This issue can be mitigated by carefully selecting the parameter range covered by the suite of simulations or by using more effective compression methods.

\subsection{Canonical Correlation Analysis (CCA)}\label{subsec:cca}

One other way to amend this flaw with $C_t$ is to perform a generalized version of PCA to modulate the effect of the sampling distribution. The first step is to project the covariance of the parameters $p_i$, $C_p$, to the data vector space. Assuming a linear model for the data vector, this is done by minimizing:
\begin{align}
    \min_{A\in \mathbb{R}^{m\times n}, b\in \mathbb{R}^m} \sum_i ((Ap_i +b) - t_i)^2, \label{eq:ptod}
\end{align}
where the matrix $A$ and vector $b$ are the parameters of the linear model. There is an exact solution to this least squared problem, given by
\begin{align}
    A &= ( TP^T - N_{\rm sim} \bar{t}\bar{p}^T)(PP^T - N_{\rm sim} \bar{p}\bar{p}^T)^{-1} \nonumber \\
    &= C_{tp}C_p^{-1},\nonumber\\
    b &= \bar{t} - A \bar{p}.\label{eq:ptod_exact}
\end{align}

In the above equations, $C_{tp}$ is the cross-covariance matrix of $t_i$ and $p_i$, $T$ and $P$ are matrices with $p_i$ and $t_i$ as columns respectively, and $\bar{t}$ and $\bar{p}$ are the mean of the data vectors $t_i$ and parameters $p_i$.

Since we assume a linear model, $A$ is the average Jacobian $\partial t/ \partial p$ esimtaed from $p_i$ and $t_i$. With this we can project all $p_i$ to data space as $\hat{t}_i = Ap_i + b$. The covariance of $\hat{t}_i$, $C_l$, is computed as follows:
\begin{align}
    C_l &= \frac{1}{N_{\rm sim}} \sum_i (\hat{t}_i - \bar{\hat{t}}_i ) (\hat{t}_i - \bar{\hat{t}}_i ) ^T\nonumber\\
    &= \frac{1}{N_{\rm sim}} \sum_i (A (p_i-\bar{p})) (A (p_i-\bar{p})) ^T\nonumber\\
    &= A \left ( \frac{1}{N_{\rm sim}} \sum_i (p_i-\bar{p})(p_i-\bar{p}) ^T \right ) A^T \nonumber \\
    &= A C_p A^T. \label{eq:cl_def}
\end{align}
As desired, $A$ directly projects the covariance of $p_i$ to the space of data vectors. For future use, note that $C_l= C_{tp} C_p^{-1} C_{pt}$. Lastly, we solve the generalized eigenvalue problem 
\begin{align}
    C_t v_i = \lambda_i C_l v_i \label{eq:kl}
\end{align}
with $C_t$ as defined on~\Cref{eq:cov}.  The eigenvalue $\lambda_i$ here encodes the variance in $t_i$ modulo the (projected) variance in $p_i$. Therefore, this identifies the highest impact parameter combinations and the directions of their effects.  We then compose the compression matrix $U$ using the components $v_i$:
\begin{align}
    U = (v_1 \ v_2 \ \cdots \ v_{n})^T \label{eq:U2},
\end{align}
and compress the data vectors using \Cref{eq:definition_compression}. In this method, unlike in PCA, the maximum number of components is automatically determined by the number of parameters $n$ varied in the model, which is typically much smaller than the length of the uncompressed data vector $m$. This is because $C_l$ is a matrix of rank $n$, as it is a projection of $C_p$, which is also a rank $n$ covariance matrix (\Cref{eq:cl_def}), and $n<m$. The eigenvalue $\lambda_i$ diverges when $v_i$ is in the non-empty kernel, $\ker(C_l)$. So the relevant $v_i$ are $v_i\in \ker (C_l)^\perp$, the orthogonal complement of $\ker (C_l)$. This problem amounts to finding the principal components of $C_t$ restricted to $\ker (C_l)^\perp = {\rm im} (C_l)$, the image of $C_l$. Assuming $A$ is rank $n$ (which should be the case for generic choices of $\bar{p}$), $\dim {\rm im} (C_l) = \dim {\rm im} (C_p) = n$ meaning there are only as many relevant $v_i$ as there are parameters. This approach was inspired by the Karhunen–Loève decomposition discussed in~\cite{dacunha2022does, raveri2020quantifying}.

Though one may motivate~\Cref{eq:kl} as an improvement to PCA as above, it is also widely known as Canonical Correlation Analysis (CCA). Given a set of two vectors such as $p_i$ and $t_i$, the goal of CCA is to identify components of $p_i$, $w\cdot p_i$, and components of $t_i$, $v\cdot t_i$, that are maximally correlated with each other. More concretely, the goal is to solve 
\begin{align}
    \max_{w\in \mathbb{R}^n, \ v\in \mathbb{R}^m } \frac{w^T C_{pt} v}{\sqrt{w^T C_p w} \sqrt{v^T C_t v}}.
\end{align}
This is equivalent to the eigenvalue problems
\begin{align}
    C_p^{-1} C_{pt} C_t^{-1} C_{tp} w &= \lambda_i w, \nonumber\\
    C_t^{-1} C_{tp} C_p^{-1} C_{pt} v = C_t^{-1}C_lv  &= \lambda_i v. \label{eq:cca}
\end{align}
The second eigenvalue problem is equivalent to~\Cref{eq:kl} in that the eigenvectors are identical. For our purposes, the $v$ vectors identify directions in data vector space that are most correlated to changes in $p_i$. CCA was previously introduced to cosmology to quantify concordance of cosmological data in~\cite{raveri2020quantifying}.

Another way to motivate and understand this approach is through mutual information. Mutual information quantifies the amount of information one measurement carries about another. It is defined as 
\begin{align}
    I(x,y) = \iint dxdy \ P(x,y) \ln \left( \frac{P(x,y)}{P(x)P(y)} \right),
\end{align}
where \( x \) and \( y \) are two random variables, \( P(x,y) \) is the joint probability distribution of \( x \) and \( y \), \( P(x) \) and \( P(y) \) are the marginal probability distributions of \( x \) and \( y \), respectively, and \( I(x,y) \) represents the mutual information, which measures the amount of information that \( x \) and \( y \) share.

In our case, suppose we are interested in how much information about the parameters is contained in some linear combination of the data vector components, $v^T t$. Furthermore, assume that the joint distribution $P(p,t)$ is a multivariate Gaussian with a covariance matrix
\begin{align}
    C = \left(\begin{matrix}
        C_p & C_{pt} \\
        C_{tp} & C_t 
    \end{matrix}\right). 
\end{align}
Then, the joint covariance matrix between $p$ and $v^T t$ is 
\begin{align}
    C^\prime = \left(\begin{matrix}
        C_p & C_{pt}v \\
        v^TC_{tp} & v^TC_t v
    \end{matrix}\right)
\end{align}
and consequently the mutual information is
\begin{align}
    I(p, v^T t) &= \frac12 \ln \left( \frac{\det C_p \det v^TC_t v}{\det C^\prime} \right) \nonumber \\
    &= \frac12 \ln \left( \frac{(\det C_p) ( v^TC_t v)}{(\det C_p ) (v^TC_t v - v^TC_{tp} C_p^{-1} C_{pt}v)} \right) \nonumber \\
    &= \frac12 \ln \left( \frac{  v^TC_t v}{ v^T(C_t  - C_{tp} C_p^{-1} C_{pt})v} \right) \nonumber \\
    &= \frac12 \ln \left( \frac{  v^TC_t v}{ v^T(C_t-C_l)v} \right).
\end{align}
Finding the vector $v$ which maximizes $I(p, v^T t)$ amounts to the following optimization problem, which corresponds to a generalized eigenvalue problem:
\begin{align}
    \max_{v} \frac{  v^TC_t v}{ v^T(C_t-C_l)v} \nonumber \\
    C_t v_i = \rho_i(C_t-C_l)v_i. \label{eq:lmim}
\end{align}
Notice that this is equivalent to the generalized eigenvalue problem in~\Cref{eq:kl}. The same set of $v_i$ solve both problems, and the eigenvalues are related as 
\begin{align}
    \lambda = \frac{\rho}{\rho - 1}
\end{align}
where $\lambda$ is the eigenvalue from~\Cref{eq:kl} and $\rho$ is the eigenvalue from~\Cref{eq:lmim}.

In summary, the following three approaches are equivalent:
\begin{itemize}
    \item Finding the principal components of $C_t$ restricted to the image of projected $C_p$
    \item Finding the components of $t_i$ most correlated with changes in $p_i$
    \item Maximizing $I(p, v^Tt)$
\end{itemize}
Going forward, this approach to DR will be referred to as CCA, but we will solve~\Cref{eq:lmim} because unlike~\Cref{eq:kl} and~\Cref{eq:cca}, matrices on both sides of the generalized eigenvalue problem are invertible which makes it more numerically stable. Furthermore, the eigenvalues are more immediately interpretable as $\exp(2\times{\rm mutual \ information})$.

\subsection{MOPED} \label{subsec:MOPED}
\begin{figure*}[t!]
    \centering
    \includegraphics[width=0.95\linewidth]{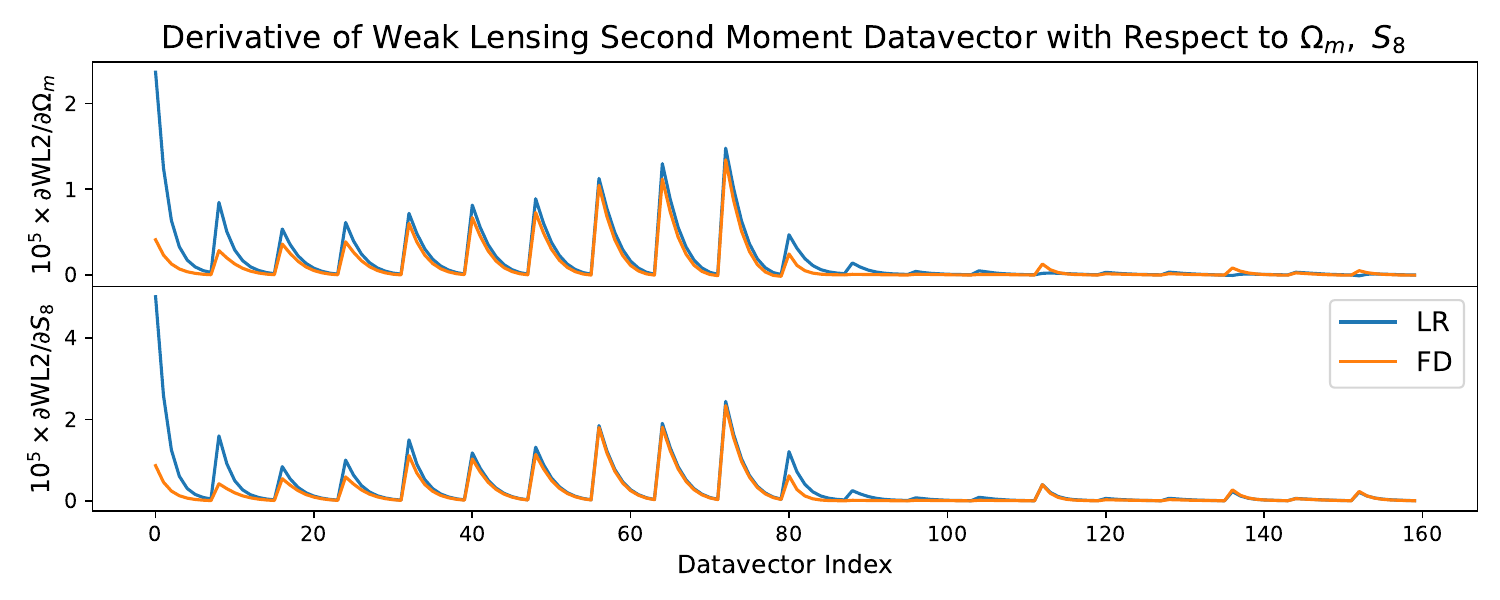}
    \vspace{-10pt}
    \caption{The derivative of the weak lensing second moments data vector with respect to $\Omega_m$ (above) and $S_8$ (below). The orange lines labeled `FD' is the derivative computed with finite differences around a fiducial cosmology. The blue lines labeled `LR' is the derivative computed by solving the linear regression problem  in~\Cref{eq:ptod}. Except for the low redshift part of the data vector (below index 20), the two methods of computing the derivative are in good agreement.  This shows that while the linear model might be a poor model, the implied derivatives are quite reasonable.}
    \vspace{-10pt}
    \label{fig:deriv_comparison}
\end{figure*}

Another common linear compression method is MOPED introduced in~\cite{heavens2000massive}. Assuming a Gaussian likelihood where the covariance matrix is independent of the parameters, the FIM is given as 
\begin{align}
    F_{\alpha\beta} &= -\left< \frac{\partial^2 \mathcal{L} }{\partial p_\alpha^{\ } \partial p_\beta^{\ }} \right> = \frac12 \Tr \left( C^{-1}_f (t_{f;\alpha}^{\ } t_{f;\beta}^T + t_{f;\beta}^{\ } t_{f;\alpha}^T) \right).
\end{align}
In the above equation, $t_{f;\alpha}$ represents the derivative of the data vector at fiducial cosmology with respect to the $\alpha$-th parameter.

Now consider the Fisher Information Matrix (FIM) of compressed data vector $b^Tt$ instead of $t$. Then, 
\begin{align}
    F_{\alpha\beta} &= \frac{ b^T (t_{f;\alpha}^{\ } t_{f;\beta}^T)b}{b^TCb}. \label{eq:moped_linear}
\end{align}
The goal of MOPED is to identify $b_\alpha$ such that $b_\alpha^Tt$ contains as much information about $p_\alpha$ as possible (maximizes $F_{\alpha\alpha}$), and $b_\alpha^Tt$ and $b_\beta^Tt$ are uncorrelated ($b_\alpha^T C_f^{\ } b_\beta^{\ } = \delta_{\alpha\beta}$). The solution to this is given below.

First, we compute $C_f$ as defined in~\Cref{eq:fidcov}. As for the derivatives, in the absence of a theoretical model for the data vector, they can be estimated using finite differences:
\begin{align}\label{eq:fdderivative} 
    t_{f;\alpha} &\equiv \overline{ \frac{\partial t_f^{\ }}{\partial p_\alpha}}  \\
    &= \frac{1}{2\delta p_\alpha}\left( \frac{1}{N_\pm} \sum_i t_{f+\delta p_\alpha,i} - \frac{1}{N_\pm} \sum_i t_{f-\delta p_\alpha,i} \right) \nonumber 
\end{align}
where \( t_{f+\delta p_\alpha,i} \) and \( t_{f-\delta p_\alpha,i} \) are data vectors at the fiducial cosmology with the \( \alpha \)th parameter shifted by \( \pm \delta{p_\alpha} \). Then, sequentially we identify vectors, $b_a$, with which to compress the data vector. 
\begin{align}
    {b}_1&=\frac{C_f^{-1} {t}_{f; 1}^{\ }}{\sqrt{{t}_{f; 1}^{T} C_f^{-1} {t}_{f, 1}^{\ }}} \nonumber \\
    b_{\alpha>1} &= \frac{C_f^{-1} {t}_{f; \alpha}^{\ } - \sum_{\beta = 1}^{\alpha-1} (t_{f;\alpha}^Tb_\beta^{\ })b_\beta^{\ } }{\sqrt{{t}_{f; \alpha}^{T} C_f^{-1} {t}_{f, \alpha}^{\ } - \sum_{\beta = 1}^{\alpha-1} (t_{f;\alpha}^Tb_\alpha^{\ })^2 }} \label{eq:MOPED}
\end{align}

The second term in the numerator and denominator of $b_{\alpha>1}$ can be understood as performing the Gram-Schmidt process with $C_f$ performing the role of the identity matrix. Assuming that the likelihood surface is Gaussian, the covariance matrix is parameter-invariant, and the parameter derivatives are  constant, $b_\alpha$ extracts all of the information relevant to inferring parameter $p_\alpha$ from the data vector. More specifically, this DR maximizes (preserves the original) Fisher Information.  The compression matrix is then the matrix whose rows are $b_\alpha$:
\begin{align}
    U = (b_1 \ b_2 \ \cdots \ b_n)^T; \label{eq:comp_mat_mop}
\end{align}
the data vectors are then compressed using  \Cref{eq:definition_compression}, as in PCA and CCA.

The main downside of MOPED is that it requires simulations at the fiducial cosmology for \( C_f \) and, when theoretical derivatives are not available, simulations with parameter shifts away from the fiducial cosmology for \( t_{f;\alpha} \). This is in  addition to the simulations at varied cosmology which one needs for the SBI. In particular, for non-Gaussian statistics, the number of simulations required for $t_{f;\alpha}$ to converge can be hard to gauge. Without proper convergence, the compressed data vectors fail to capture relevant information. There are ways to reduce this computational load by using proxies such as Gaussian simulations which are less expensive to generate. However, this may prove suboptimal, especially for non-Gaussian statistics.

\subsection{e-MOPED}

One way of overcoming this computational downside of MOPED is to attempt to infer $C_f$ and $t_{f;\alpha}$ from the simulations at varied cosmologies. The starting point here is once again to assume a linear model for the data vectors. This is quite natural given that MOPED already assumes the parameter derivatives to be constant. We solve~\Cref{eq:ptod} again with~\Cref{eq:ptod_exact}. Then, by assumption, the columns of $A$ are $t_{f;\alpha}^{\ }$. As seen in~\Cref{fig:deriv_comparison}, derivatives computed this way are comparable to the derivatives computed with finite differences. Furthermore, if we assume the model to be linear, we can `shift' $t_i$ from $p_i$ to $\bar{p}$, an identical point in parameter space,
\begin{align}
    t_i \mapsto \tilde{t}_{f,i} = t_i - A(p_i-\bar{p}),
\end{align}
getting rid of the linear effects of being away from $\bar{p}$. Then we can estimate $C_f$ at $\bar{p}$ with these shifted $\tilde{t}_{f,i}$. We will call this $\tilde{C}_f$ which is equivalent to 
\begin{align}
    \tilde{C}_f &= \frac{1}{N_{\rm sim}}\sum_i ( \tilde{t}_{f,i} -\bar{\tilde{t}}_{f} )(\tilde{t}_{f,i}-\bar{\tilde{t}}_{f})^T \nonumber \\
    &=\frac{1}{N_{\rm sim}} \sum_i (t_i - A(p_i-\bar{p}) - \bar{t}) (t_i - A(p_i-\bar{p})- \bar{t})^T\nonumber\\
    &= C_t + C_l \nonumber \\
    & \  - \frac{1}{N_{\rm sim}} \sum_i  A(p_i-\bar{p})(t_i -\bar{t})^T  + (t_i -\bar{t})(p_i-\bar{p})^TA^T \nonumber \\ 
    &= C_t + C_l - ( AC_{pt} + C_{tp}A^T) =  C_t - C_l. 
\end{align}
Indeed, $\tilde{C}_f$ takes all the variance in $C_l$ out of $C_t$, since the variance encoded in $C_l$ (the projection of $C_p$) is the variance of parameters. $\tilde{C}_f = C_f$ if noise is entirely parameter independent and the model is truly linear. 


With these estimates of $\tilde{C}_f$ and $\tilde{t}_{f;\alpha}$ we proceed with~\Cref{eq:MOPED} and ~\Cref{eq:comp_mat_mop} to form the compression matrix and compress the data vectors. We will call this approach e-MOPED, short for `easy'-MOPED. Though these estimates of $C_f$ and $t_{f;\alpha}$ are imperfect and imprecise, we will show that the performance of e-MOPED is comparable to MOPED but with drastically reduced computational requirements.




\subsection{It's All Optimization}

All of these linear DR techniques can be framed as optimization problems~\cite{ghojogh2019eigenvalue}:
\begin{align}\label{eq:optimization}
    {\rm PCA: }\ \ &\max_{v} \frac{v^T C_t v}{v^Tv} \nonumber\\
    {\rm PCA{\text -}f: }\ \ &\max_{v} \frac{v^T C_f v}{v^Tv} \nonumber\\
    {\rm MOPED: }\ \ &\max_{v} \frac{v^T A_fA_f^T v\footnotemark }{v^TC_fv} \nonumber\\
    {\rm e{\text -}MOPED: }\ \ &\max_{v} \frac{v^T AA^T v}{v^T(C_t-C_l)v} \nonumber\\
    {\rm CCA: }\ \ &    \max_{w\in \mathbb{R}^n, \ v\in \mathbb{R}^m } \frac{w^T {\rm cov}(t, p) v}{\sqrt{w^T C_p w} \sqrt{v^T C v}} \\
    {\rm reduces \ to }  \ &\max_{v} \frac{v^T C_t v}{v^T C_l v} \ {\rm equivalently} \  \frac{v^T C_t v}{v^T (C_t-C_l) v}  \nonumber
\end{align}
\addtocounter{footnote}{-1}
\footnotetext{Technically, as seen in~\cite{heavens2000massive}, MOPED sequentially maximizes $v^T(t_{f;i}^{\ }t_{f;i}^T)v/v^TC_f^{\ }v$ then performs Gram-Schmidt on the identified vectors. However, this achieves the same effect.}
All of these reduce to generalized eigenvalue problems where the matrix on the left hand side is the matrix in the numerator, and the matrix on the right hand side is the matrix in the denominator. The solution maximizes the numerator while ortho-normalizing with respect to the bilinear form in the denominator. 

Firstly, note that PCA variants' loss functions have units of $t^2$. This means that they are not invariant under linear transformations of the data vectors, and are therefore sensitive to the order of magnitudes of data vector components which is irrelevant to the information content. Their other shortcomings were discussed in detail in~\Cref{subsec:PCA}. For these reasons, prior to empirical tests, it is realistic to expect the PCA variants to perform worse compared to the other linear methods. 

MOPED and e-MOPED's loss functions have units of $p^{-2}$. They are not invariant under linear transformations of the parameters, but that is not an issue. The definition of the parameters of interest are entirely within our control, e.g. $\sigma_8$ versus $S_8$. The difference between e-MOPED and MOPED is the use of $\tilde{C}_f = C_t-C_l$ as opposed to $C_f$, and $A$ as opposed to $A_f$. It may be argued that this is more appropriate for dimensionality regulation for SBI, since this relaxes the assumption that the covariance matrix and Jacobian are constant throughout parameter space. By using an ``average'' covariance matrix and Jacobian, it is a more realistic attempt at maximizing the Fisher Information across  parameter space. 
Furthermore, MOPED diagonalizes $C_f$ but cannot diagonalize $C_t$ meaning MOPED projects $t_i$ so that the entries are correlated. 
On the other hand, e-MOPED diagonalizes $C_t-C_l$ which is dominated by $C_t$ meaning e-MOPED projects $t_i$ so each entry adds new information. This is shown in~\Cref{fig:cov}.

CCA and e-MOPED use the same denominator but different numerators. The same denominator $v^T(C_t-C_l)v$ indicates that both minimize noise uncorrelated with changes in parameter. Both numerators seek to highlight the directions in data space most informative to parameter inference. CCA is concerned with mutual information, which is a global quantity over all $p_i$, so it achieves this by maximizing data-parameter covariance without needing derivatives.  
On the other hand, e-MOPED uses $(p_i, t_i)$ pairs across parameter space to estimate a $\tilde{C}_f$ and $A$ at $\bar{p}$ to maximize Fisher information locally at $\bar{p}$, as a proxy for maximizing Fisher information everywhere. So it follows the gradient of parameters in data space at $\bar{p}$, which by the inverse function theorem is aligned with $\partial t/\partial p$. So, it  explicitly aligns $v$ with the rows of $A$.

With all this in mind, there are three main questions this paper attempts to answer.
\begin{enumerate}
    \item How do CCA or e-MOPED compare to MOPED in the SBI framework given that they need drastically fewer simulated data vectors?
    \item Can these linear DR methods outperform non-linear methods?
    \item Which of these should we use for a given problem?
\end{enumerate}

\section{Non-linear Methods}\label{Sec:non-linear_compressors}

Recently, there have been approaches to DR using neural networks with various loss functions. The theoretical upside to the non-linear methods is that with the flexibility and expressiveness of a neural network, heavily non-linear relationships between parameters and data vector or highly non-Gaussian distribution of simulated data vectors may be disentangled. 
For all of the following non-linear methods, the compression is done with a neural network $f_\phi$ such that $c = f_\phi(t)$ is the compressed data vector. 

\subsection{Neural Network with MSE} \label{subsec:MSE}

One natural approach to using neural networks for DR in SBI is driven by the insight that the most helpful thing to know for parameter inference may just be the parameters themselves. It is possible to train a neural network to decode cosmological parameters from simulated data vectors. Let $f_\phi : \mathbb{R}^{m} \rightarrow \mathbb{R}^n$ be a neural network with parameters $\phi$. Then we solve 
\begin{align}
    \min_{\phi} \sum_i (f_\phi(t_i) - p_i)^2. \label{eq:nnmse}
\end{align}
We will call this approach NN-MSE as it is a neural network (NN) with a mean squared error (MSE) loss function \citep{Jeffrey2021,jeffrey2024dark,gatti_wph1,gatti_wph2}.\footnote{A linearized version of NN-MSE where $f(t_i) = A t_i + b$ was tried, but all other methods presented greatly outperformed this approach and therefore was left out of the text. } 

{How well should we expect this to perform? Assuming that a well trained $f_\phi$ is an unbiased estimator of $p_i$ from $t_i$, the (co)variance of the estimate has a lower bound per the Cramer-Rao (CR) bound. Assuming Gaussian error, we have:
\begin{align}
    f_\phi (t_i) &= p_i + \epsilon_i  \\
    \epsilon_i &\sim \mathcal{N}(\vec{0}, C(p_i) ) \ \textrm{and}\ C(p_i) \geq (F(p_i))^{-1}
\end{align} where $C(p_i)$ is the covariance of the estimator at $p_i$, $F(p_i)$ is the FIM  at $p_i$, and the matrix inequality $A\geq B$ implies $A-B$ is positive semidefinite. With $f_\phi (t)$ as the data vectors, the covariance of a posterior estimate ($C_p$) is bounded from below by the inverse of the FIM of the output of $f_\phi$. Assuming the parameter dependence of $C(p)$ is negligible, the FIM of $f_\phi(t) = p + \epsilon$ is equal to $C^{-1}$ since the `data'-parameter Jacobian is the identity; $\partial f_\phi(t)/\partial p = I$. In total, $C_p \geq C \geq F^{-1}$ in a chain of CR bounds. The philosophy behind NN-MSE is that $f_\phi(p)$ should perfectly capture the information content implied by the FIM.}


{However, there is no guarantee of this, indeed it is quite unlikely that either CR bound comes close to saturation. Furthermore, the non-linearity of $f_\phi$ introduces variance that is an artifact of $f_\phi$ and irrelevant to posterior estimation. Compressing the data by decoding the parameter value is an interpretable and human idea. However, it is not clear that parameter estimation is the best way to extract information relevant to posterior estimation. This is what motivates the approach presented in the next section - training a neural network to pick out features of the data most useful to a posterior estimator. }

Practically speaking, since the way that the parameters affect the data vectors are non-trivially different it is most effective to train separate neural networks for each parameter. We use Optuna~\cite{akiba2019optuna} to optimize the network structure hyperparameters for each parameter. Let $f_{\alpha}$ be the neural network trained to predict the $\alpha^{\rm th}$ parameter. Then, the compressed data vector is 
\begin{align}
    c = f_\phi(t) = \left( f_1(t) \ f_2(t) \ \cdots \ f_n(t)
    \right)^T
\end{align}
which is a non-linear approximation of the parameters as inferred from the data vector $t$ via neural networks.

\subsection{Variational Mutual Information Maximization} \label{subsec:VMIM}

Another approach is Variational Mutual Information Maximization (VMIM) introduced in~\cite{barber2004algorithm,Jeffrey2021}. The idea is to maximize the mutual information between parameters and compressed data vectors, where the data vectors are compressed with some neural network ($f_\phi$) as above. 
\begin{align}
I({f_\phi(t)}, {p}) & =D_{\mathrm{KL}}(P({f_\phi(t)}, {p}) | P({f_\phi(t)}) P({p})) \nonumber\\
& = \mathbb{E}_{P({f_\phi(t)}, {p})}[\log P({p} | {f_\phi(t)})] - H({p})\nonumber\\
&\supset \sum_{i} \log P({p_i} | {f_\phi(t_i)}) \label{eq:vmim_loss}
\end{align}
where the sum iterates over $p_i, \ t_i$
.
Since we do not know the exact posterior $P({p_i} | {f_\phi(t_i)})$, we must model the posterior with a neural density estimator with parameters $\varphi$: $\tilde{P}({p_i} | {f_\phi(t_i)}; \varphi)$. It is known that modeling the posterior with a NDE gives a lower bound of the loss function;
\be \sum_{i} \log P({p_i} | {f_\phi(t_i)}) \geq \sum_{i} \log \tilde{P}({p_i} | {f_\phi(t_i)}; \varphi ).\ee
The latter serves as the ``variational mutual information'' as it is the variational (NDE) approximation to the mutual information. So in total, we vary $\phi$ and $\varphi$ together to solve 
\begin{align}
    \max_{\phi, \varphi} \sum_{i} \log \tilde{P}({p_i} | {f_\phi(t_i)}; \varphi)
\end{align}
$\varphi$ and the trained NDE is ultimately irrelevant to the DR. The neural network component, $f_\phi(t)$ defines the compressed data vector. 

Note that this is a generalization of CCA. As discussed in~\Cref{subsec:cca}, CCA can be understood as maximizing mutual information assuming a Gaussian distribution for $P(p,t)$ and restricting the DR to linearity. The use of an NDE to approximate the posterior relaxes the Gaussian approximation, and the use of a neural network $f_\phi$ relaxes the linearity.

\subsection{Information Maximizing Neural Networks} \label{subsec:IMNN}

Information Maximizing Neural Networks (IMNN) is another approach to DR recently introduced in~\cite{charnock2018automatic,Makinen2024}. In essence, this is a non-linear generalization of MOPED. Instead of considering the FIM of $b^Tt$ as in~\Cref{eq:moped_linear}, we consider that of $f_\phi(t)$ which is a data vector $t$ compressed with a neural network $f_\phi$.
\begin{align}
    F_{\alpha\beta} &= \overline{(f_\phi(t))_{;\alpha}}^TC_{f_\phi}^{-1}\overline{(f_\phi(t))_{;\beta}} \nonumber \\
    C_{f_\phi} &= \overline{(f_\phi(t) - \overline{f_\phi(t)})(f_\phi(t) - \overline{f_\phi(t)})^t} 
\end{align}
where the average parameter derivatives and $C_{f_\phi}$ are computed at a fixed value of parameters. The derivatives are computed with finite differences of the outputs of neural networks. IMNN finds $f_\phi$ that best compresses the data vectors by maximizing the determinant of $F_{\alpha\beta}$ at various points in parameter space with some regularizing terms. 

\subsection{Practicality}

Under certain assumptions, these loss functions, MSE, variational mutual information, and Fisher information, are all related. Assuming a Gaussian posterior, the pertinent term in~\Cref{eq:vmim_loss} is just the $\chi^2$~\cite{sui2023evaluating}:
\begin{align}
    I(p, f_\phi(t)) &\supset \sum_i \log P(p_i|f_\phi(t_i))  \\
    &\supset \sum_i -\frac12 (p_i-f_\phi(t_i))^T C_p^{-1}(p_i-f_\phi(t_i))\nonumber
\end{align}
which is akin to the MSE loss with weights assigned according to the posterior covariance $C_p$. The link between mutual information and Fisher information is often invoked to put a bound or estimate on the former using the latter in the neuroscience community. In the statistical context most relevant to our use case where the noise is Gaussian and non-negligible, one can invoke the Cramer-Rao bound to show~\cite{brunel1998mutual,wei2016mutual}:
\begin{align}
    I(p, f_\phi(t)) - H(p) \geq -\int d^n p \ P(p) \frac{1}{2} \ln \left(\frac{(2 \pi e)^n}{\operatorname{det} (F(p))}\right)
\end{align}
where $\operatorname{det} (F(p))$ is the determinant of the FIM conveyed by $f_\phi(t)$ at parameter space location $p$. This is to say, training a neural network to increase the FIM determinant is akin to increasing the lower bound of $I(p, f_\phi(t))$. IMNN indirectly accomplishes the optimization posed by VMIM, which is in turn similar to MSE optimization under realistic assumptions. 

Pragmatically speaking, we were only able to obtain reliable results with VMIM when we pretrain with MSE loss and add a MSE loss term to the original VMIM loss.  However, even in the best case scenario we find that the results of VMIM to be $\sim30\%$ worse than results of NN-MSE in terms of the Figure of Merit (FoM) defined in~\Cref{subsec:SBI}. Training IMNN requires the estimate of derivatives and covariances at various points in parameter space with an extremely large number of simulated data vectors which we find impractical to obtain for non-Gaussian summary statistics.

Given these pragmatic considerations and the mathematical relationships established above, we opt to only test NN-MSE as a benchmark for how well non-linear methods perform compared to linear methods. 

\begin{table}[]
    \centering
    \begin{tabular}{ |  >{\centering\arraybackslash}p{0.11\textwidth} |  >{\centering\arraybackslash}p{0.22\textwidth} | >{\centering\arraybackslash}p{0.13\textwidth} | } 
    \hline
    Abbreviation & Name & \# of Sims Used\\ \hline\hline
    NN-MSE & Neural Network-Mean Squared Error & 9124\\ \hline
    PCA & Principal Component Analysis & 9124  \\ \hline
    PCA-f & Principal Component Analysis at Fiducial & 6400 \\ \hline
    CCA & Canonical Correlation Analysis & 9124 \\ \hline
    MOPED &  Massively Optimized Parameter Estimation & 54280  \\ \hline
    e-MOPED & easy-MOPED & 9124 \\ \hline
    \end{tabular}
    \caption{The abbreviations, full names, and number of simulations used for the 6 DR methods mainly discussed in the rest of the paper. NN-MSE is the only non-linear method, rest are linear compression. MOPED uses 6400 for covariance estimation and around 3000 for each parameter derivative estimation.}
    \vspace{-15pt}
    \label{tab:abrv_table}
\end{table}

\section{Simulations, Statistics, and SBI} \label{Sec:data}
\subsection{Simulations} \label{subsec:sims}

\begin{table}
\centering
\begin{tabular}{c c c}
\toprule
\textbf{Parameter} & \textbf{Mocks parameters} & \textbf{Analysis prior} \\
 & \textbf{distribution} & \\
\midrule
$\Omega_{\textrm{m}}$ & mixed active-learning  & $ \mathcal{U}(0.15,0.52)$ \\ & in $\mathcal{U}(0.15,0.52)$   \\ 
$S_8$ & mixed active-learning  & $\mathcal{U}(0.5,1.0)$ \\& in $\mathcal{U}(0.5,1.0)$ \\
\midrule
$w$ & $\mathcal{N}(-1,\frac{1}{3})$ for $-1<w<-\frac{1}{3}$ & $\mathcal{U}(-1,-\frac{1}{3})$ \\
 & $0$ ${\textrm{else}}^{*}$  \\
$n_s$ & $\mathcal{N}(0.9649, 0.0063)$   \\
$h$ & $\mathcal{N}(0.7022, 0.0245)$ \\
$\Omega_{\textrm{b}}h^2 $&  $ \mathcal{N}(0.02237, 0.00015)$\\
$m_{\nu} $& $ \exp(\mathcal{U}[\log(0.06), \log(0.14)])$ \\
\midrule
$A_{\rm IA}$ & $\mathcal{U}[-3, 3]$   &  $\mathcal{U}[-3, 3]$ \\
$\eta_{\rm IA}$ & $\mathcal{U}[-5, 5]$ &  $\mathcal{U}[-5, 5]$ \\
$m_{1}$ & $\mathcal{N}(-0.0063,0.0091)$ \\
$m_{2}$ & $\mathcal{N}( -0.0198,0.0078)$ \\
$m_{3}$ & $\mathcal{N}( -0.0241,0.0076)$ \\
$m_{4}$ & $\mathcal{N}(-0.0369, 0.0076)$ \\
$\bar{n}_i(z)$ & $p_{\textsc{HyperRank}}(\bar{n}_i(z) | x_{\textrm{phot}})$  \\ 
\bottomrule
\end{tabular}
\caption {Model parameters (first column), their distribution in the Gower Street simulations and in the mock catalogs derived from these simulations (second column), and the prior used in the SBI analysis (third column). The analysis prior can differ from the distribution of the samples as long as these parameters have been explicitly used during the training of the Neural Density Estimators (NDEs) when learning the likelihood surface. For parameters where the analysis prior is not indicated, it means the likelihood dependence on that parameter is not explicitly learned, and the parameter is effectively marginalized over according to the mocks parameters distribution.  $^*$In our simulation runs we usually excluded values of $w$ less than $-1$, but 64 simulations were run without this constraint. These were still used to train our NDEs, although we applied a strict prior of $w > -1$ for the analysis.} 
\label{parameter}
\end{table}

For this work, we utilize the weak lensing mock maps developed for the simulation-based inference analyses of the Dark Energy Survey Year 3 (DES Y3) data \citep{jeffrey2024dark, gatti_wph1, gatti_wph2}. These mock maps are based on the Gower Street simulation suite \citep{jeffrey2024dark}, which includes 791 $w$CDM gravity-only full-sky N-body simulations. 
These simulations, produced using the \textsc{PKDGRAV3} code \citep{potter2017pkdgrav3}, vary the cosmological parameters $\Omega_{\textrm{m}}$, $\sigma_8$, $n_s$, $h (\equiv h_{100})$, $\Omega_{\textrm{b}}$, $w$, $m_{\nu}$ (see~\Cref{parameter}). Each full-sky simulation is used to create 25,312 pseudo-independent noisy convergence DES Y3 mock maps. The map-making process is detailed in \citep{gatti_wph1, gatti_wph2}. In summary, we create 4 noisy mock maps for each full-sky simulation, corresponding to the 4 DES Y3 tomographic redshift bins \citep{y3-sompz,y3-sourcewz}. The maps are created in the format of \texttt{HEALPIX} maps \citep{GORSKI2005} with \texttt{NSIDE = 512}. These mocks feature realistic shape noise \citep{y3-shapecatalog} and include source clustering effects \citep{source_clustering}. The mock-making procedure involves several free parameters: four multiplicative shear biases $m_i$ (one for each tomographic bin), four redshift distributions $n_i(z)$ (one for each tomographic bin), and parameters $A_{\textrm{IA}}$ and $\eta_{\textrm{IA}}$ that control the amplitude and redshift evolution of intrinsic alignment. For each DES Y3 mock catalogue, we randomly draw one of these parameters from their prior; for the redshift distributions, we draw from one of the multiple realizations provided by \cite{y3-sompz}, which account for the uncertainties in the redshift calibration of the DES Y3 $n(z)$.

The MOPED compression method requires a covariance matrix and derivatives for our summary statistics, estimated through finite differences. For this purpose, we also use simulations from the \texttt{CosmoGridV1} suite \citep{cosmogrid1}. From this suite, we selected a set of 200 full-sky simulations at the fiducial cosmology with $\sigma_8 = 0.84$, $\Omega_{\textrm{m}}=0.26$, $w=-1$, $h=0.6736$, $\Omega_{\textrm{b}}=0.0493$, and $n_{\textrm{s}}=0.9649$, as well as 50 pairs of simulations where only one of the parameters is varied at a time. We used these full-sky simulations to create 4000 pseudo-independent mocks at the fiducial cosmology to estimate the covariance matrix and 2000 pairs of mocks for each of the parameters varied in the analysis to estimate the derivatives. The simulations (and summary statistics) at fiducial cosmology are also used as our ``target'' data vector when producing posteriors in the reminder of the paper.\footnote{$m_\nu$ derivatives for \texttt{WPH} did not converge, so was not useful for MOPED. For fairness, we don't consider variation in $m_\nu$ for all methods.}

For each of our simulations and weak lensing mocks, we also create ACT DR6-like Cosmic Microwave Background (CMB) lensing maps. This is done following \citep{omori2024}; we first generate a noiseless lensing map at redshift \( z=3 \) using the density field produced by the N-body simulation and assuming the Born approximation. Then, the power spectrum of the lensing field at \( z > 3 \) is computed, from which a Gaussian realization is generated and added to produce the full CMB lensing map. We then add uniform Gaussian noise according to \citep{CMBL_ACT6}, and the ACT DR6 mask is applied.

\subsection{Summary Statistics} 
\label{subsec:sumstats}

\begin{table}[t!]
    \centering
    \begin{tabular}{ |  >{\centering\arraybackslash}p{0.13\textwidth} |  >{\centering\arraybackslash}p{0.25\textwidth}|  >{\centering\arraybackslash}p{0.06\textwidth} | } 
    \hline
    Name & Meaning & Length\\ \hline\hline
    \wls & Weak lensing second moments & 160 \\ \hline
    \texttt{WL3} & Weak lensing third moments & 160 \\ \hline
    \texttt{WL23} & \wls\ and \texttt{WL3}  & 320 \\ \hline
    \texttt{WPH} &  Wavelet phase harmonics & 256 \\ \hline
    \texttt{CMBWL} & CMB lensing $\times$ Weak lensing second moments &32 \\  \hline
    \all & All of the above & 608 \\ \hline
    \end{tabular}
    \caption{The names, meanings, and lengths of the data vectors discussed throughout the paper. \wls\ and \texttt{CMBWL} are Gaussian. \texttt{WL3} and  \texttt{CMBWL} are non-Gaussian.}
    \vspace{-10pt}
    \label{tab:dv_table}
\end{table}

The summary statistics used in this work are as follows: weak lensing second and third moments, weak lensing wavelet phase harmonics, and weak lensing - CMB lensing second moments. Second moments probe the variance of the field and are generally considered part of the class of Gaussian statistics, as they would be sufficient to describe the statistical properties of the field if it were Gaussian. Third moments and wavelet phase harmonics, on the other hand, are non-Gaussian, as they can also capture non-Gaussian features of the field. Specifically, third moments probe the skewness of the field, while wavelet phase harmonics are even more general. These statistics are applied to 'smoothed' variants of the weak lensing and CMB lensing maps, with the choice of smoothing method varying according to the specific statistic employed: moments use top hat filters, while wavelet phase harmonics and the scattering transform use wavelet filters \citep{Cohen1995, Mallat1999, VDB1999}.

\begin{figure*}[t!]
    \vspace{-10pt}
    \hspace{-20pt}\includegraphics[width=0.36\textwidth]{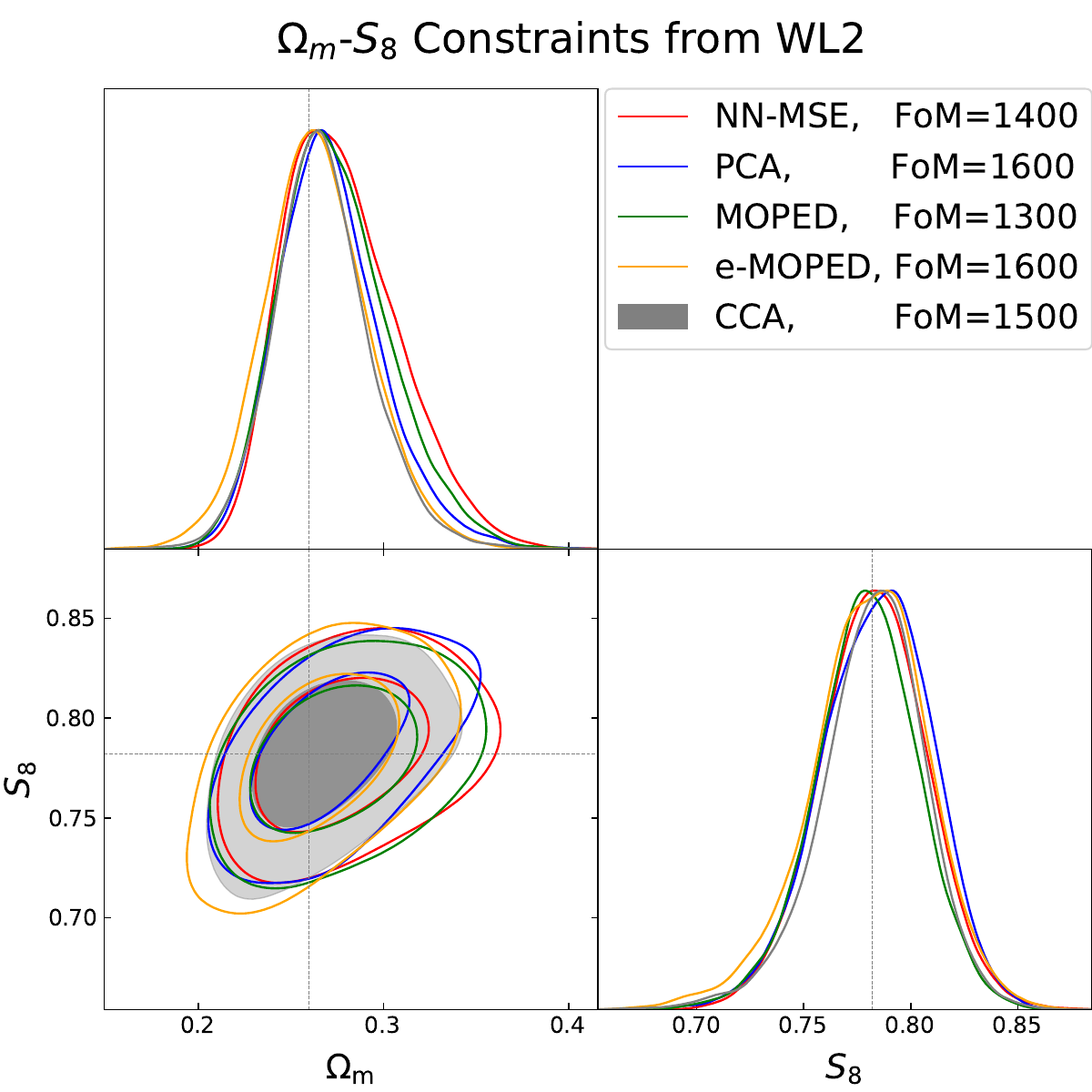}
    \includegraphics[width=0.36\textwidth]{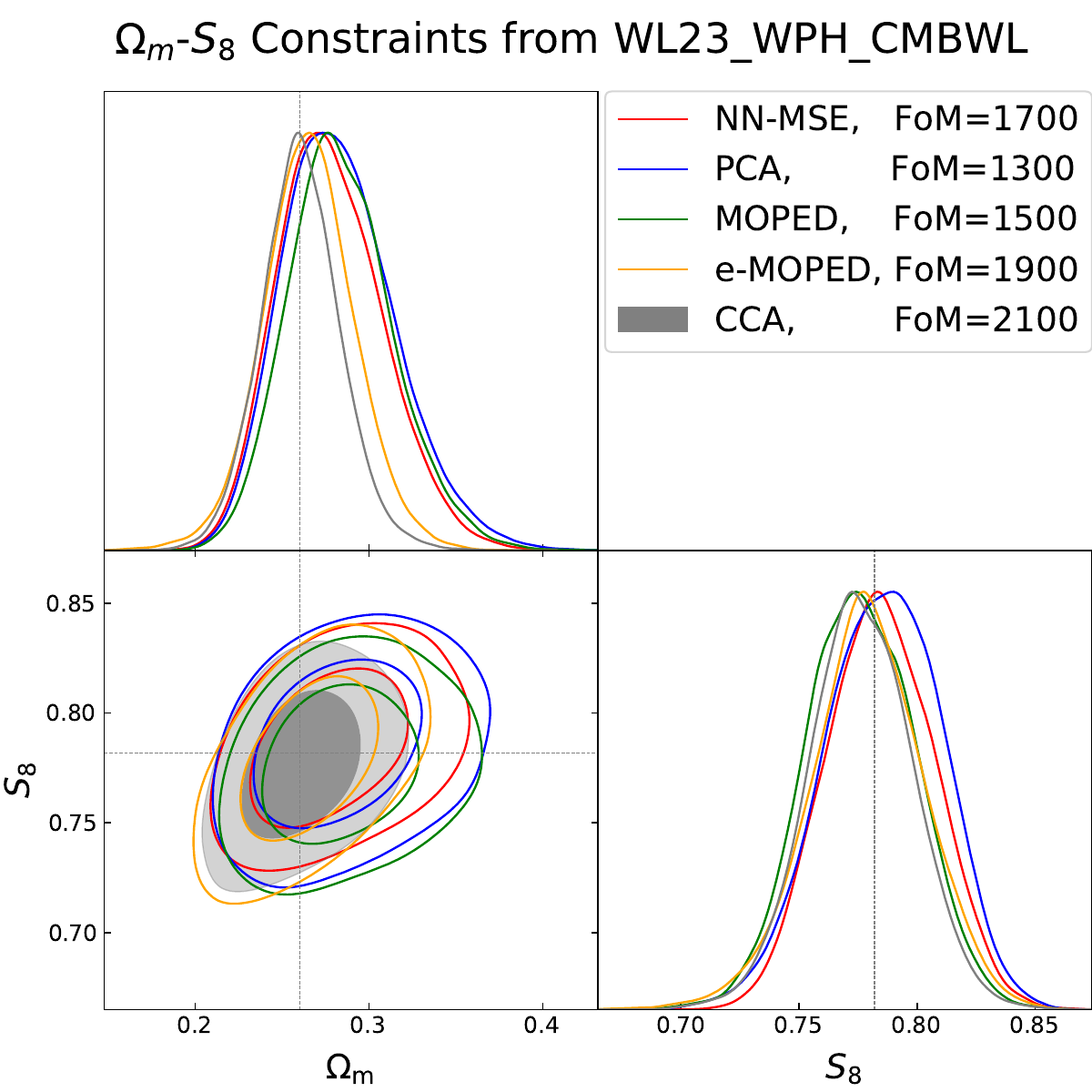}
    \raisebox{0.1\height}{\includegraphics[width=0.3\textwidth]{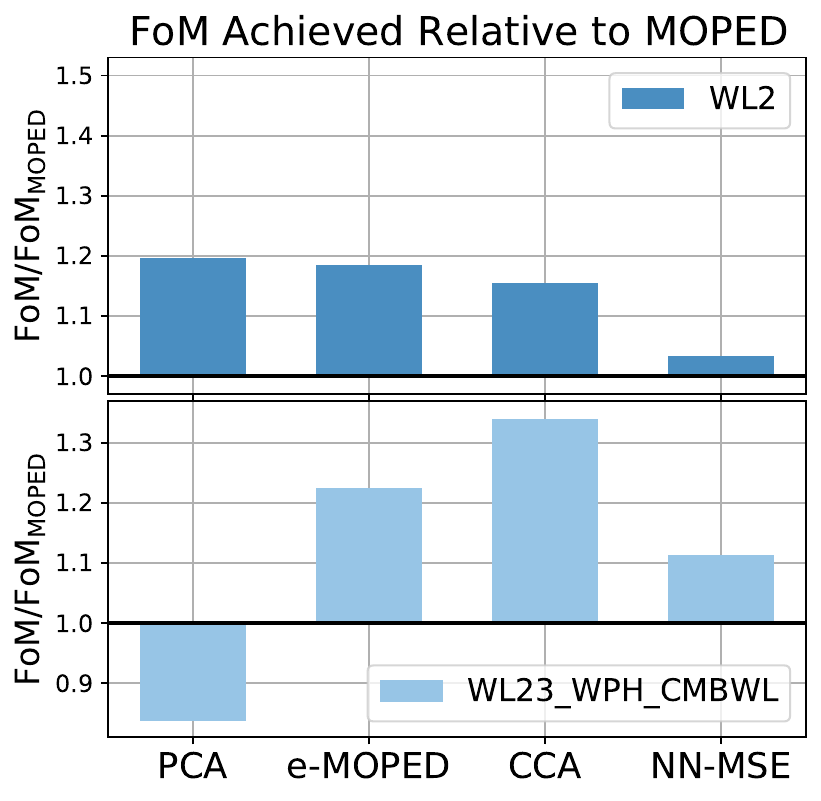}}\hspace{-20pt}
    \vspace{-10pt}
    \caption{Constraints on $\Omega_m-S_8$ from the WL second moments  (\wls; left) and all the summary statistics considered  in our study (\all; middle). The data vectors are compressed using five different approaches: NN-MSE (red), PCA (blue), MOPED (green), e-MOPED (yellow), and CCA (gray). The FoM presented are means computed over $20$ NDE realizations and rounded to the nearest $100$. The contours are chosen at random from the NDE realizations. The right panel is a bar chart showing the ratios between the FoM achieved and the FoM achieved using MOPED for \wls\ (above) and \all\ (below). 
    CCA and e-MOPED consistently outperform both MOPED and NN-MSE. PCA has the best performance when only considering \wls\ but when non-Gaussian statistics are included the performance drops drastically, because PCA is ill-suited to extract information from low-SNR statistics such as \texttt{WPH}.  }
    \vspace{-8pt}
    \label{fig:which_is_best}
\end{figure*}

For the second and third moments of the weak lensing maps \citep{VanWaerbeke2013,Petri2015,Vicinanza2016,Chang2018,Peel2018,Vicinanza2018, G20,moments2021}, we first smooth the maps using top-hat filters, and we consider eight smoothing scales $\theta_0$ equally (logarithmically) spaced from $8.2$ to $221$ arcmin. The second and third moments estimators are:
\begin{align}
\avg{\hat{\kappa}^2_{\theta_0}}(i, j) &= \Avg_p \left( \kappa_{\theta_0,p}^i \, \kappa_{\theta_0,p}^j \right) \nonumber \\
\avg{\hat{\kappa}^3_{\theta_0}}(i, j, k) &= \Avg_p \left( \kappa_{\theta_0,p}^i \, \kappa_{\theta_0,p}^j \, \kappa_{\theta_0, p}^k \right),
\end{align}
where $\kappa_{\theta_0,p}^i$ is the smoothed lensing mass map of tomographic bin $i$ ($i,j,k$ refer to different tomographic bins), and the average is over all pixels $p$ on the full sky. In the remainder of the paper, we will use \texttt{WL2} and \texttt{WL3} to indicate the second and third moments data vectors, and \texttt{WL23} to indicate when the two are concatenated. With our choice of redshift bins and smoothing scales, \wls\ and \texttt{WL3} data vectors are 160 entries long each. 

For the weak lensing - CMB lensing cross second moments, these are defined as:
\begin{equation}
\avg{{\hat{\kappa}_{\theta_0}\hat{\kappa^{\rm CMB}}}_{\theta_0}}(i, j) = \Avg_p \left( \kappa_{\theta_0,p}^i \, \kappa_{\theta_0,p}^j \right),
\end{equation}
where $\kappa^{\rm CMB}_{\theta_0}$ is the simulated smoothed ACT DR6 CMB lensing map. Before smoothing the CMB lensing and weak lensing maps, we applied to both of them the intersection of the DES and ACT DR6 masks. In the remainder of the paper, we will indicate the weak lensing - CMB lensing second moments are \texttt{CMBWL}. With our choices, \texttt{CMBWL} data vector is 32 entries long.

Wavelet phase harmonics (WPH) correspond to the second moments of smoothed weak lensing mass maps that have undergone a non-linear transformation \citep{Cohen1995, Mallat1999, VDB1999, Allys2020}. To achieve this smoothing, a directional, multi-scale wavelet transform is applied to the maps. Consider a smoothed map \( \kappa_{{n,\ell}}^i(\vec{\theta}) \), where \( n \) indicates the size of the filter (approximately \( 2^{n+1} \) pixels), and \( \ell \) denotes the orientation. The WPH statistics utilized in this work are:
\begin{align}
    S01(i, j, n) &= \Avg_p \Avg_{\ell} \left( |\kappa_{{n,\ell}}^i| \,\, \kappa_{{n,\ell}}^j \right) \nonumber \\
    C01{\delta \ell 0}(i, j, n_1, n_2) &=  \Avg_p \Avg_{\ell} \left( |\kappa_{{n_1,\ell}}^i| \,\, \kappa_{{n_2,\ell}}^j \right)  \nonumber\\
    C01{\delta \ell 1}(i, j, n_1, n_2) &= \Avg_p \Avg_{\ell} \left( |\kappa_{{n_1,\ell+1}}^i| \,\, \kappa_{{n_2,\ell}}^j \right) 
\end{align}
where $n_1<n_2$. We average over all pixels \( p \) and also over the three values \( 0, 1, 2 \) of the rotation index \( \ell \) (corresponding to the three possible orientations of the directional wavelet). The number \( n \) varies from \( 0 \) to \( 5 \). These statistics are a subset of those used in \citep{gatti_wph1, gatti_wph2}, selected because they primarily probe the non-Gaussian features of the fields. In the remainder of the paper, we will refer to all of these wavelet statistics collectively as \texttt{WPH}. With our choices of redshift bins, $\ell$, and $n$, the \texttt{WPH} data vector is 256 entries long. 

Lastly, \texttt{WL23\_WPH\_CMBWL} refers to the combination of all three of these summary statistics. In total, this data vector is 608 entries long.

\subsection{Posterior Estimation (SBI)} \label{subsec:SBI}

We estimate our posteriors using simulation-based inference (SBI). Specifically, we use neural density estimators (NDE's) to learn the likelihood surface from our noisy mock measurements of the compressed summary statistics \( c_i \). In particular, we employ Masked Autoregressive Flows (MAF; \cite{Papamakarios2017}), as implemented in the package pyDELFI \citep{Alsing2018}, to estimate the conditional distribution \( p(c_i | p_i) \). NDE's approximate this distribution with estimates \( q(c_i | p_i; \phi') \), where the network parameters \( \phi' \) are optimized by minimizing a loss function. 

The density estimation is performed using a subset of the parameter set \( \theta = \left[ \Omega_{\textrm{m}}, S_8, w, A_{\textrm{IA}}, \eta_{\textrm{IA}} \right] \) and our compressed data vectors. This approach implies that the remaining parameters, which vary in the mock productions, are effectively marginalized over \citep{momentnets} according to the prior distributions applied when sampling these parameters during the mock generations.
\begin{figure*}[t!]
    \vspace{-12pt}
    \includegraphics[width=0.95\textwidth]{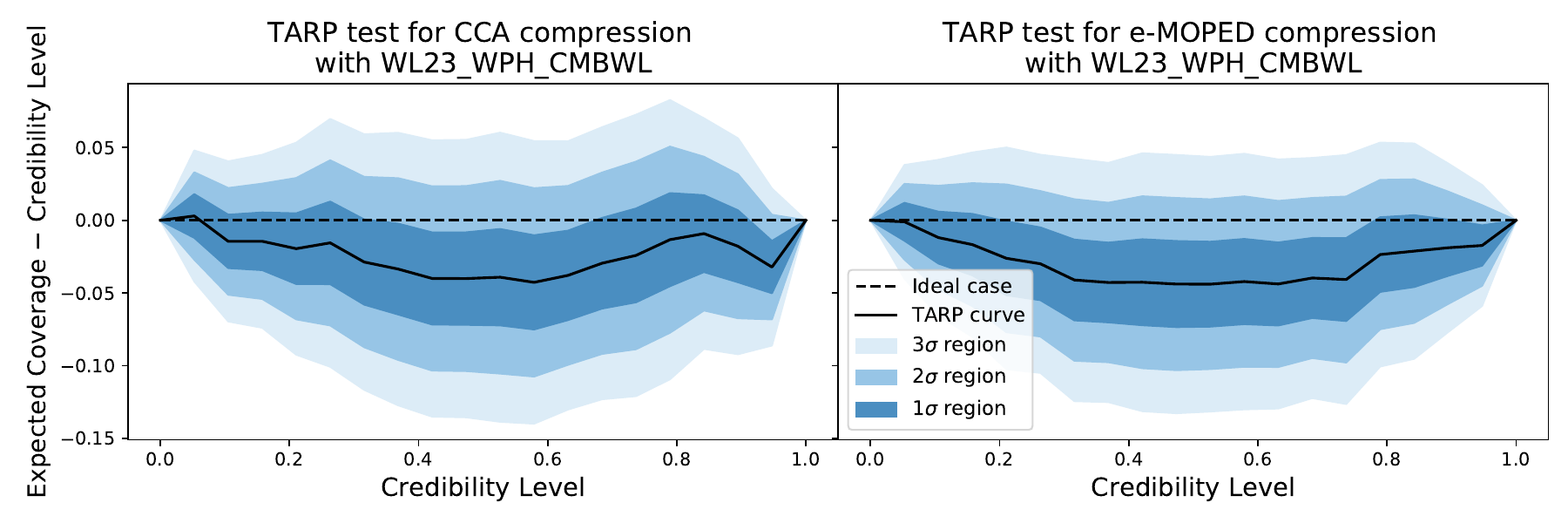}
    \vspace{-12pt}
    \caption{The TARP coverage test result and confidence intervals  for posteriors estimated with all summary statistics (\all) compressed with CCA (left) and e-MOPED (right). We choose to plot Expected Coverage$-$Credibility Level, instead of Expected Coverage which is more common, for visibility. The black solid line is the result of the TARP test, the black dotted line is the ideal case, and the shaded regions are 1, 2, and 3$\sigma$ uncertainty bands. There is no indication of over or under-confidence. The curve always being below the ideal is indicative of bias, but the significance is less than $2\sigma$ in both plots. Note that this does not mean the posteriors themselves are $1\sim2\sigma$ off the truth (\Cref{fig:which_is_best} shows no evidence of bias in the posteriors), but that there is a less than $2\sigma$ indication of \textit{some} level of bias in the estimated posteriors.}
    \vspace{-10pt}
    \label{fig:tarp_single}
\end{figure*}

After learning the likelihood surface, we compute the likelihood at ``observed'' data $t_{\textrm{obs}}$. We use the average of the compressed summary statistics measured in the \texttt{CosmoGridV1} simulations at fiducial cosmology as our observed data. The final posteriors are derived via Markov chain Monte Carlo (MCMC) sampling of the likelihood, while considering wide flat priors  for $\Omega_{\textrm{m}}, S_8, w, A_{\textrm{IA}}, \eta_{\textrm{IA}}$. This MCMC sampling is carried out using the publicly available software package \texttt{EMCEE} \citep{Foreman-Mackey2013}, an ensemble sampler with affine-invariant properties designed for MCMC. 

We evaluate each compression method for their performance using the FoM metric, defined as
\begin{align}
    {\rm FoM} = \left(\det  \text{Cov}(\Omega_m,S_8)\right)^{-1/2}
\end{align}
where $\text{Cov}(\Omega_m,S_8)$ is the posterior covariance of $\Omega_m$ and $S_8$. We observed that the NDE training process introduces a small variance in the final FoM (around 5\%). To address this, we repeated the training 20 times for each study case and averaged the final chains, resulting in more stable FoM values. 

\section{Results} \label{sec:results}
\subsection{Which is `Best'?} \label{subsec:ranking}

We used PCA, NN-MSE, MOPED, e-MOPED, and CCA to compress \wls\ and \all\ to constrain $\Omega_m$ and $S_8$. ~\Cref{fig:which_is_best} shows the resulting contours and FoM. The \wls\ contours show how the compression methods handle the simplest case of a Gaussian statistic with the highest signal to noise ratio (SNR). The \all\ contours show how the compression methods handle a much longer data vector and non-Gaussian statistics with lower SNR. Note that the standard deviation of the mean is $\sim$20 which is much smaller than the typical differences between pairs of FoM figures. 

This answers the questions posed at the end of~\Cref{Sec:linear_compressors}. CCA and e-MOPED consistently outperform NN-MSE and MOPED, even with a longer data vector and non-Gaussian statistics. Conversely, this means that NN-MSE and MOPED, which are often used, are not truly lossless compressors. Additionally, CCA and e-MOPED greatly reduce the number of simulated data vectors required while not sacrificing performance over MOPED. For concreteness, in our experiments MOPED requires an order of magnitude more simulated data vectors (see~\Cref{tab:abrv_table}), which is extremely costly to acquire with N-body simulations. CCA and e-MOPED are also much easier to implement with less tuning required. All the posteriors shown are unbiased, so we compare the methods purely based on FoM. 

PCA performs very well with \wls\ but performs poorly with \all. This is because \wls\ is a Gaussian statistic with high SNR. That means the variance due to change in parameters dominates $C_t$ over noise contributions. Therefore, the principal components of $C_t$ limited to \wls\ are effective at extracting information relevant to the cosmological parameters. However, PCA fails with \all\ because the noisy part of the data vector, \texttt{WL3} and \texttt{WPH}, dominates over shifts in cosmological parameters. CCA and e-MOPED should be preferred over PCA in general. 

The performance of all other methods improves when going from \wls\ to \all, which is the expected behavior when adding more information to compress. However, the improvement with MOPED is very marginal so MOPED falls drastically behind other linear methods. This is because the derivatives of non-Gaussian statistics computed with finite differences do not converge easily. It is possible that with even more simulations dedicated to the finite difference computation, MOPED compression of \all\ performs better. However, as stated above, that is computationally costly.

{ A toy exercise indicates that CCA and e-MOPED are more effective than NN-MSE when the response to parameter is linear within the simulation suite prior. The details of the toy exercise can be found in~\Cref{app:toy}. The results here indicate that \all\ considered here falls within that regime of linearity.}

We also use the Tests of Accuracy with Random Points (TARP) coverage test to assess whether our posteriors with CCA or e-MOPED are robust. TARP is a diagnostic for overconfident, under-confident, or biased posterior estimators. We use the TARP package with chains generated from our NDE's to find TARP curves, as well as the included bootstrapping method to put error bars on the curves. TARP curve being above then falling below the ideal indicates over-confidence, and below then above the ideal indicates under-confidence. TARP curve staying below the ideal indicates biased posteriors.~\cite{lemos2023sampling}


\Cref{fig:tarp_single} shows the results of the TARP test on posteriors estimated with \all\ compressed with CCA and e-MOPED. Neither of the tests indicate issues with confidence. The curves can be seen staying below the ideal, but the ideal and TARP curves are at most $2\sigma$ away. This does not mean posteriors are $2\sigma$ away from truth. It indicates that there is less than $2\sigma$ significance for \textit{some} bias in the posteriors, likely dominated by prior volume effects near the edges of our $p_i$ ranges since the posteriors shown in~\Cref{fig:which_is_best} are all fairly centered. Thus we may conclude that CCA and e-MOPED compression do not result in issues with confidence or bias when posteriors are estimated.

While CCA and e-MOPED approach DR with different philosophies, their performance is similar and superior to other methods tested here. We have also shown that they do not cause confidence or bias issues in the final posteriors. Since they are easy to implement and quick to run, the best practice for DR before SBI may be to try both.

\subsection{Is Everything Useful?} \label{subsec:components}
\begin{figure}[t!]
    \centering
    \vspace{-10pt}
    \includegraphics[width=0.4\textwidth]{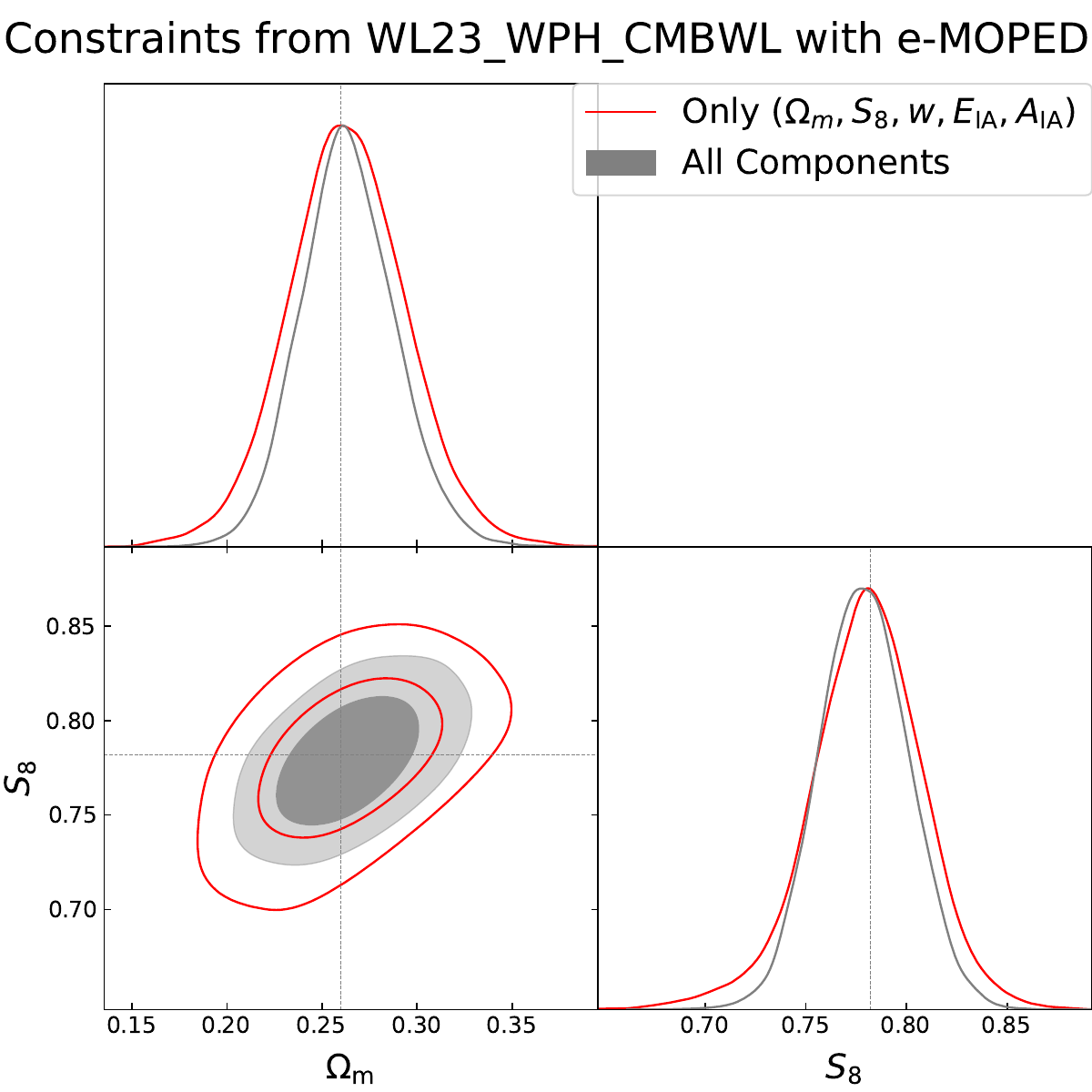}
    \vspace{-10pt}
    \caption{Constraints on $\Omega_m-S_8$ based on \all\ using e-MOPED. Posteriors were generated only using the five compressed data vector components corresponding to $(\Omega_m, S_8, w, \eta_{\rm IA}, A_{\rm IA})$ (red),  and using all 16 components (gray). Contours indicate that including information about other parameters tightens constraints on $\Omega_m-S_8$. Across all compression methods, using only 5 most important components results in 20$\sim$50\% smaller FoM.}
    \vspace{-10pt}
    \label{fig:needmore}
\end{figure}

Let $c$ refer to the entire compressed data vector. Truncate the data vector to define $c_a$ and $c_b$ such that $\vec{c} = (\vec{c_a}, \vec{c_b})$. The posterior obtained on parameter $\theta$ just considering $c_a$ is 
\begin{align}
    P(\theta|c_a) &= \frac{P(c_a|\theta)P(\theta)}{P(c_a)} = \frac{P(\theta)}{P(c_a)} \int d c_b P(c_a, c_b|\theta)  \nonumber\\\nonumber
    &= \frac{P(\theta)}{P(c_a)} \int d c_b \frac{P(\theta|c_a, c_b)P(c_a,c_b)}{P(\theta)}\\ 
    &= \int dc_b P(\theta|c_a,c_b)P(c_b|c_a). 
\end{align}  
On the other hand, posterior obtained on $\theta$ considering the entire data vector $t$ is 
\begin{align}
    P(\theta|c_a, c_b) &= \int d \tilde{c}_b P(\theta|c_a,\tilde{c}_b)\delta(c_b-\tilde{c}_b). 
\end{align} 
So, unless $P(\theta|c_a,c_b)$ is independent of $c_b$ or $P(c_b|c_a)$ is sharply peaked at the observed value of $c_b$, $P(\theta|c_a, c_b)$ is going to be a sharper distribution than $P(\theta|c_a)$. In the former case, consider the case that $c_b$ is just a Gaussian variable independent of $\theta$ or $c_a$. For the latter, consider the case that $c_b$ is just some linear multiple of $c_a$. In both of these cases, it is evident that including $c_b$ on top of $c_a$ would not contribute to the constraining power on $\theta$.  

All that is to say, it is useful to have more data than not. This is widely accepted and applied - e.g. one achieves a tighter constraint on $\Omega_m$ by combining CMB data with weak lensing data. However, in the case of these compressed data vectors it is not always intuitive. For MOPED and NN-MSE, elements of the compressed data vector correspond to specific parameters. So, one might think that to constrain $\Omega_m$ and $S_8$, it should be sufficient to use the parts of the compressed data vector corresponding to $\Omega_m$ and $S_8$ (`$c_a$' above). However, the above and our experiments below show that in fact the other parts of the data vector (`$c_b$' corresponding to $\Omega_b, n_s, h, dm_i, dz_i$) are useful in constraining $\Omega_m$ and $S_8$.

\begin{figure*}[t!]
    \centering
    \vspace{-10pt}
    \includegraphics[width=0.4\textwidth]{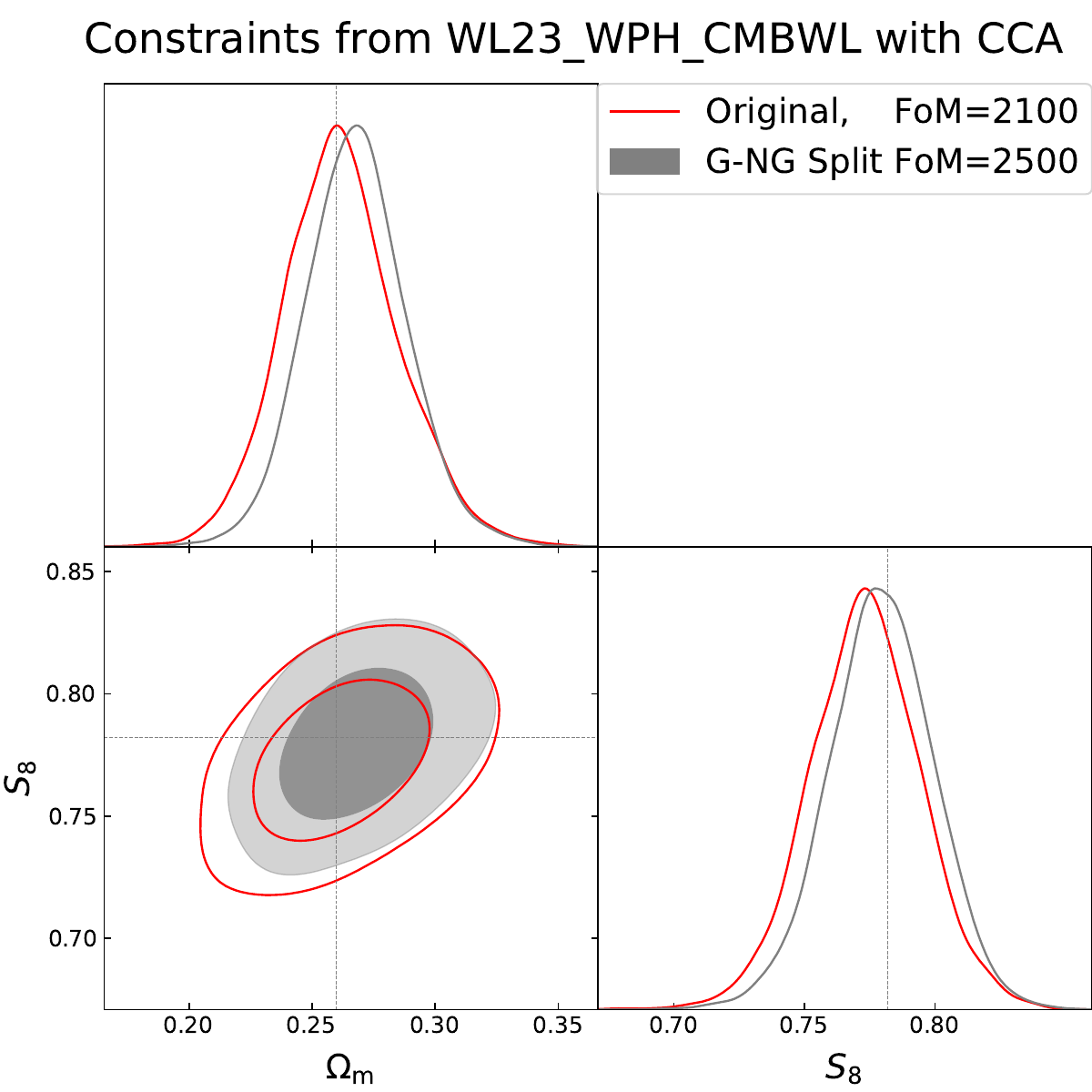}
    \includegraphics[width=0.4\textwidth]{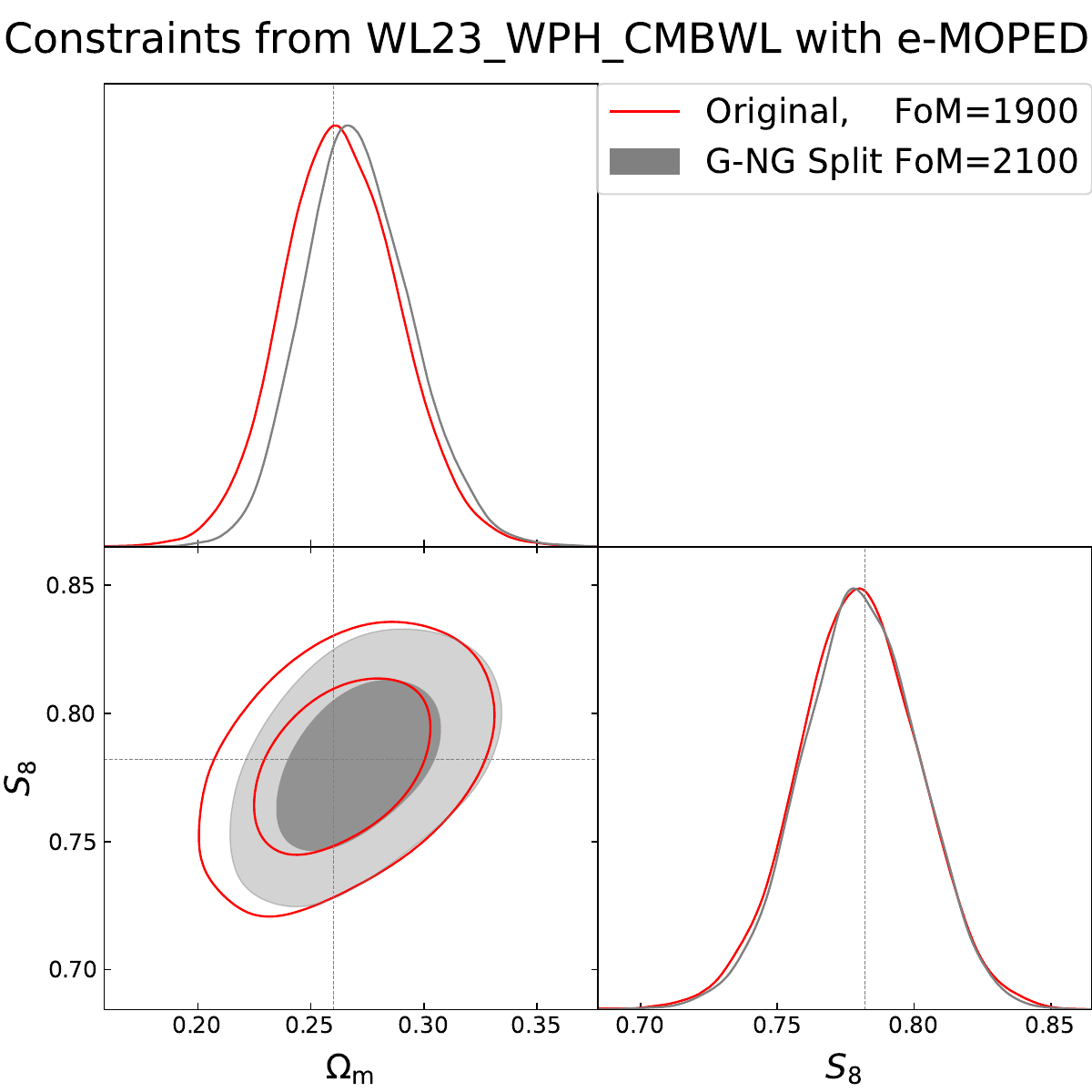}
    \vspace{-10pt}
    \caption{Constraints on $\Omega_m-S_8$ based on \all, data vectors compressed using CCA (left) and e-MOPED (right). The red contours are generated by compressing the Gaussian (\wls, \texttt{CMBWL}) and non-Gaussian (\texttt{WL3}, \texttt{WPH}) statistics together. The gray contours are generated by compressing them separately and concatenating for a total compressed data vector length of 32.  There is a significant increase in constraining power by splitting Gaussian and non-Gaussian parts of the data vector to compress them separately. }
    \label{fig:doubleup}
\end{figure*}
\Cref{fig:needmore}~ shows that with e-MOPED, indeed the FoM is significantly better when all 16 components, instead of just a small subset, of the compressed data vector are used to constrain just the two parameters of relevance. We make this point with e-MOPED in this figure as the parameter to compressed component correspondance is intuitive. However, the point stands more generally with all other compression methods, including CCA. The compressed components of CCA are ordered in descending order of correlation with the parameters. It is very helpful to have all 16 components, as opposed to just the top 5. Only using the top 5 components results in a 20$\sim$50\% smaller FoM across all data vector and compression method combinations.

One practical concern with this is that NDE's may not be able to handle high dimensional data vectors. Suppose our NDE's can only handle up to $N$ components but there are more than $N$ parameters. Then the resulting compressed data vectors would be too long for our NDE's. CCA orders the compressed data vector components according to their correlation to parameters. The choice would be to pick the top $N$  most correlated components. MOPED, e-MOPED, and NN-MSE compressed components each have clear correspondences to parameters. Then we must make physics-informed decisions about which components to keep. For example, if $N = 14$, it would make more sense for us to discard $\bar{n}_1, \ \bar{n}_2$ components than $\eta_{\rm IA}, \ A_{\rm IA}$ components as we know intrinsic alignment is very degenerate with $\Omega_m, \ S_8$.

Following this recommendation, all experiments on this paper compress data vectors to 16 dimensions with all methods considered. That is, for MOPED, e-MOPED, and NN-MSE, all 16 components corresponding to the 16 parameters are included. For CCA, all 16 components are included. For PCA (and PCA-f) the 16 components with the highest variance are included. 

\vspace{-5pt}

\subsection{PCA with Varied Cosmology} \label{subsec:pca_global}

As discussed in~\Cref{subsec:PCA}, PCA-f has theoretical shortcomings in the SBI framework in that the principal components are those of pure noise, where as the principal components of PCA has a chance of picking up on the variation on cosmological parameters. We compare the constraints on $\Omega_m$ and $S_8$ while using PCA and PCA-f to compress \wls\ in~\Cref{fig:pcaf} of~\Cref{app:pca-f}. Indeed, PCA-f performs much worse compared to PCA. This is not just true for the case of \wls\ shown here, but also true for any combination of the datasets considered. Though we argue above that one should use CCA or e-MOPED for DR before SBI, if one is to use PCA it should be PCA of $C_t$ not $C_f$ at fixed cosmology.

\subsection{Compress Separately Gaussian and non-Gaussian Statistics}\label{subsec:separately}

Gaussian statistic and non-Gaussian statistics have vastly different SNR and have different relationships to the parameters of interest. This makes it potentially problematic to compress Gaussian and non-Gaussian statistics together. For example, CCA finds linear combinations of data vector entries maximally correlated with corresponding linear combinations of parameters. These optimal parameter combinations are different between \wls\ and \texttt{WL3}. So, when you compress \wls\ and \texttt{WL3} together, the compression does not extract information from  \wls\ or \texttt{WL3} as efficiently as when they are compressed separately. Since Gaussian statistics are much more constraining than non-Gaussian statistics alone, it is important to make sure that Gaussian information is extracted optimally. Furthermore, since all Gaussian statistics are functions of the matter power spectrum in some form or another, their parameter dependencies are similar. It stands to reason then, that it might be best to compress Gaussian statistics separately for optimal Gaussian information extraction then concatenate the compressed non-Gaussian data vectors. We try precisely this idea with CCA and e-MOPED for \all\ in~\Cref{fig:doubleup}.

Using CCA we compress the Gaussian part of the data (\wls\ and \texttt{CMBWL}) and the non-Guassian part of the data (\texttt{WL3} and \texttt{WPH}) separately. This results in 16 entries for compressed Gaussian data and another 16 for compressed non-Gaussian data, making the final compressed data vector 32 entries long. We can see that for both e-MOPED and CCA, there is a significant increase in constraining power by compressing Gaussian and non-Gaussian data vectors separately. 
\begin{figure*}[t!]
    \vspace{-7pt}
    \includegraphics[width=0.95\textwidth]{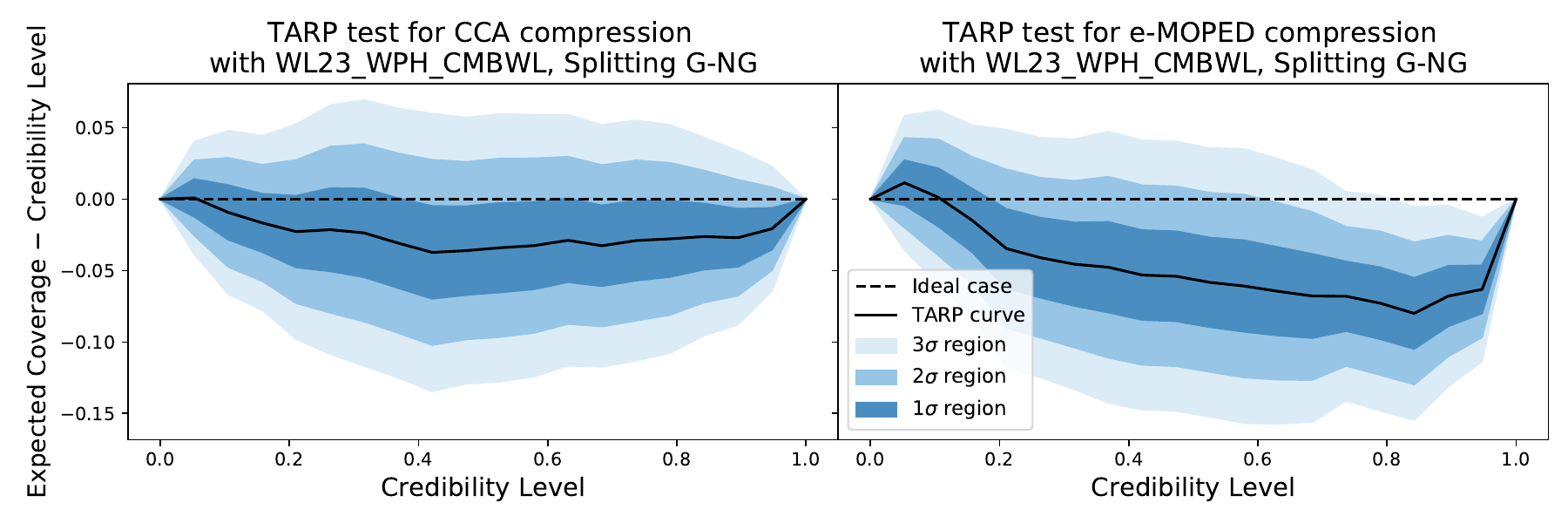}
    \vspace{-12pt}
    \caption{The TARP coverage test result and error bars for posteriors estimated with \all's Gaussian and non-Gaussian parts compressed separately by CCA (left) and e-MOPED (right). We choose to plot Expected Coverage$-$Credibility Level, instead of Expected Coverage which is more common, for visibility. Black solid line is the result of the TARP test, the black dotted line is the ideal case, and the shades are 1, 2, and 3$\sigma$ regions. Again, there is no indication of over or under-confidence. However, on the right the evidence for bias is stronger than in~\Cref{fig:tarp_single} where the Gaussian and non-Gaussian statistics were compressed together. This shows that one should be careful when choosing to compress Gaussian and non-Gaussian statistics separately. }
    \vspace{-12pt}
    \label{fig:tarp_double}
\end{figure*}

In~\Cref{fig:tarp_double} we use the TARP coverage test to examine any issues with posteriors generated with this technique. Again, we find no significant indication of over or under-confidence as the TARP curves mostly stay under the ideal. With e-MOPED the TARP curve is $3\sigma$ away from the ideal at certain points, while with CCA the situation is similar to that in~\Cref{fig:tarp_single}. We conclude that this splitting between Gaussian and non-Gaussian greatly boosts the FoM using CCA without introducing bias, but it poses a risk of biased results when using e-MOPED. This result is however specific to our data vector size, the number of parameters varied, and the number of simulations at our disposal to train the NDE's. 

This technique calls for a great deal of caution and one must ask several important questions before applying it. For example:
\begin{itemize}
    \item Are there other ways to split the data vector? For instance, rather than dividing the data vector by Gaussian and non-Gaussian statistics, should the splitting be guided primarily by the SNR?
    \item Would compressing different parts of the data vector separately increase the overall length of the data vector? If there are already 30 parameters and correspondingly CCA results in data vectors in $\mathbb{R}^{30}$, then using this technique results in data vectors in $\mathbb{R}^{60}$ which would be unwieldy for NDE's. 
\end{itemize}

In the latter, the question is how to balance the benefits reported in this section and those reported in~\Cref{subsec:components}. If there are too many parameters, it may be better to just compress the Gaussian and non-Gaussian statistics together. Or it may be better to compress separately, but only use a part of each of the compressed data vectors. There are no clear a priori heuristic for deciding the procedure and the solution depends on the specifics of the application.

\section{Conclusion} \label{Sec:conclusion}

\subsection{Summary}
We have studied a set of linear and non-linear compression methods for cosmological inference problems. The linear methods are PCA, CCA, MOPED and a variation proposed here, e-MOPED (see \S\ref{Sec:linear_compressors}). The nolinear methods are based on neural networks and are NN-MSE, VMIM and IMNN (\S\ref{Sec:non-linear_compressors}). 
We provide theoretical arguments and experiments with  simulations to identify the best methods and practices when performing DR. Our quantitative results are in the context of  SBI for cosmological parameter inference though some of our findings are applicable for covariance matrix based MCMC analysis as well. We use tomographic weak lensing cosmological analysis as our application and consider a number of summary statistics: second and third moments and a set of wavelet based statistics, Wavelet Phase Harmonics and Scattering Transforms. We include the leading sources of systematic uncertainty and consider combinations of galaxy and CMB WL as well. 

For DR, CCA is very well motivated as a linear technique maximizing mutual information by projecting the data vectors to subspaces most correlated with changes in parameters, without using derivatives. MOPED is a technique that maximizes Fisher information at some  fidcucial parameter values $p_f$. e-MOPED is our proposal for using samples across parameter space to maximize Fisher information at $\bar{p}$ as a proxy for Fisher information everywhere by computing the average Jacobian and projecting data vectors along the parameter gradient. This has the added benefit of greatly reducing the computational cost. For the non-linear methods, a combination of theoretical bounds and practicality suggests that NN-MSE is preferable over VMIM and IMNN for our use case. 

\begin{figure*}[t!]
    \centering
    \includegraphics[width=0.99\linewidth]{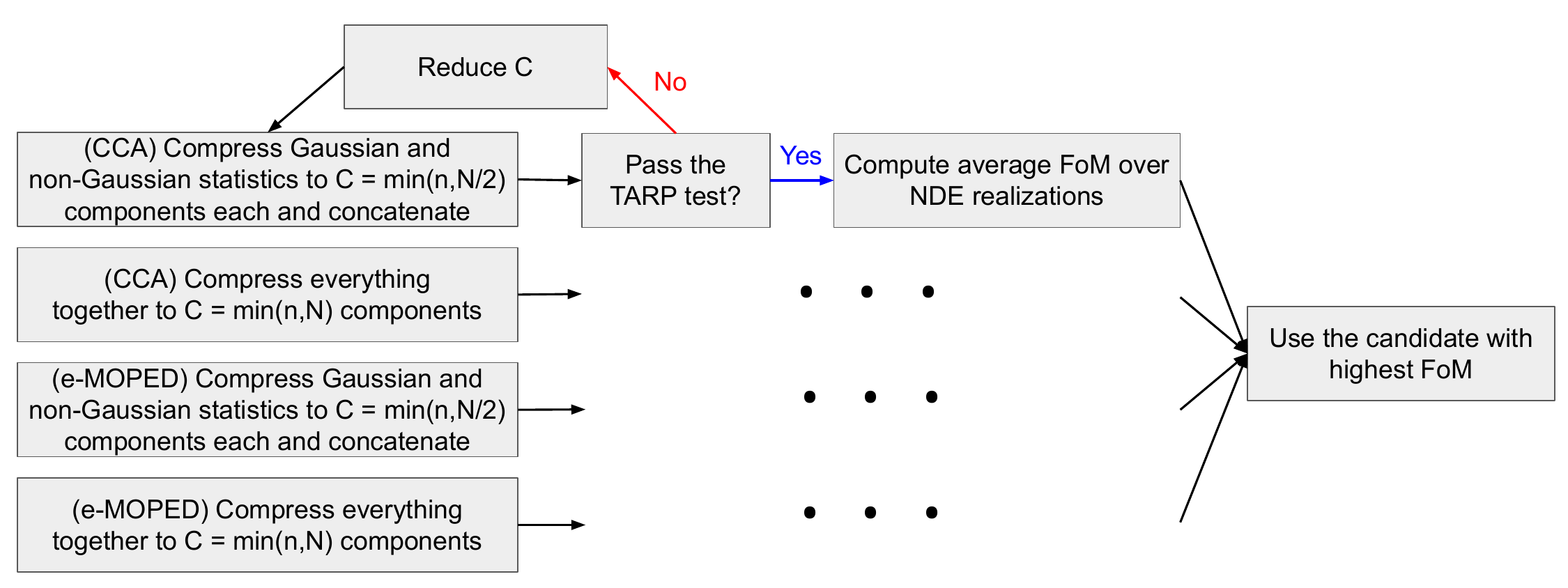}
    \caption{Flowchart on how to decide on which compression scheme to use. $n$ is the total number of parameters, $N$ is the upper bound of data vector length that NDE's can handle. We find $N$ to be $\sim30$. The four options on the left are all valid options, so it is important to test all of them for their possible flaws with TARP and get a reliable estimate of the FoM from multiple NDE realizations. One should try to tune the data vector length until all four options pass the TARP test, and use the compression that results in the highest average FoM.}
    \label{fig:flowchart}
\end{figure*}

Figure \ref{fig:which_is_best} summarizes the key findings in our tests with simulated data vectors. When implemented appropriately, as described in \S\ref{sec:results}, none of the methods tested result in bias, so we judge them purely based on FoM.  CCA and e-MOPED consistently outperform not just all other linear methods, but also NN-MSE given our setup. This also means we identify information loss in MOPED and NN-MSE compressions. For the task of SBI for cosmological parameter inference, CCA and e-MOPED are therefore our recommended DR methods. While PCA works well in some situations, it does poorly with low SNR data vectors, and there are {a priori} reasons to believe PCA and PCA-f would be unsuitable for SBI. {MOPED is severely limited by the difficulty of numerical derivative convergence. Toy exercise explored in~\Cref{app:toy} reveal that the relevant problem is in a regime where NN-MSE is inadequate. }

{At the same time, given the non-linear relationship between our parameters and data vectors, it is very plausible that there is a neural network architecture and loss function that out-perform CCA and e-MOPED. We find VMIM and IMNN to be impractical, and NN-MSE to underperform, but there may be a neural DR technique that is practical and powerful. This is especially the case for levels of noise and non-linearity beyond those considered here. }

We also find that components of the compressed data vector ($c$) not corresponding to the primary parameters of interest, $\Omega_m$ and $S_8$, are quite useful in further constraining these two parameters. If at all correlated with $\Omega_m$ and $S_8$, additional components of $c$ can add useful information. Only using parts of $c$ naively relevant to constraining the parameters of interest can result in $20\% \sim 50\%$ loss in FoM. Furthermore, it is often  beneficial to compress Gaussian and non-Gaussian statistics separately and concatenate the compressed data vectors for SBI. Both of these recommendations result in longer compressed data vectors. This resulted in a $\sim20\%$ increase in FoM in the cases tested. The benefits of these recommendations must be carefully balanced with the robustness of NDE's with longer data vectors in mind.

{We leave the following open questions for future investigation: Under what circumstances (or with which summary statistics) do the assumptions of e-MOPED or CCA break down enough that other methods are preferable? Is there a neural compression algorithm that would out-perform e-MOPED or CCA? How do the conclusions here translate to future surveys?}

\subsection{Recommendations for SBI}

\Cref{fig:flowchart} condenses the recommendations to a flowchart. For any compression before SBI there are two choices to make: CCA or e-MOPED, and compressing Gaussian and non-Gaussian statistics together or separately. In a scenario with $n$ parameters and NDE's capable of handling maximum data vector length $N$, the latter choice becomes very pertinent if $n>N/2$. In our experiments, we find $N\sim30$. Of the four resulting compression schemes, there is no way to know a priori which will perform the best. So we recommend trying all four and tuning the length of the compressed data vector used for SBI until all four schemes pass the TARP test. Then, compute the average FoM for each of the four schemes by averaging over NDE realizations. Finally, the compression scheme with the highest average FoM should be selected. In our case, without any tuning, we can see that compressing Gaussian and non-Gaussian statistics separately with CCA passes the TARP test and results in the highest FoM of $2500$. Therefore, we conclude that SBI of $\Omega_m$-$S_8$ using \all\ should use this compression scheme.

Our experiments show that following the above recommendations can significantly increase FoM over practices common in the literature, while providing insights into the differences between Gaussian and non-Gaussian statistics. With careful inspection and implementation of DR techniques, advances can be made in both cosmological parameter constraints, and our understanding of the relationship between parameters and non-Gaussian statistics. 

\subsection{Outlook for Different Datasets and Analysis Frameworks}

We have focused our study on weak lensing summary statistics, particularly through a full simulation-based inference approach. In cosmology, studies of large-scale structure can utilize various observables and frameworks. Alternative frameworks often rely on:
\begin{itemize}
    \item Fully analytical models for the data vector and covariance, assuming a Gaussian likelihood (e.g., \citep{kids2pt,giblin2021kids,PhysRevD.105.023520, PhysRevD.105.023514})
    \item Theory models for the data vector with simulation-based covariance, using a Gaussian likelihood (e.g., \citep{G20, gatti2022dark})  
    \item Simulation-based models for the data vector (obtained, e.g., using emulators) with simulation-based covariance, using a Gaussian likelihood (e.g., \citep{harnois2021cosmic,Zuercher2022,Zuercher2024})
\end{itemize}

Analyses that rely on simulations for the covariance (and possibly for the models of the observables) can benefit from the data reduction techniques presented here. The exceptions are analyses where analytical models for the data vector and covariance are fully trusted (such as the primary fluctuations of the CMB temperature and polarization, which are well described by analytical covariance matrices and are not relevant to this work).   A representative example of a LSS observable is the WL third moment (\texttt{WL3}) or shear 3-point function for which a theoretical model exists but  analytical models for the covariance matrix may not be accurate or realistic. One could significantly reduce the number of simulations required for covariance matrix estimation by first using CCA or e-MOPED. These techniques, contrary to MOPED, rely only on estimates of the data vectors at different points in parameter space, which can be provided by the theory models. With a much smaller compressed data vector, it would  then be possible to run a more reasonable number of simulations at a fixed cosmology to estimate the covariance matrix. While our results are based on mock catalogs for a Stage 3 lensing survey, we expect the improvements we have found to hold generally though further work is needed to make quantitative statements for Stage 4 surveys and observables beyond WL. 

Other observables that can benefit from data reduction techniques include three-dimensional galaxy redshift surveys, secondary anisotropies of the CMB, the Lyman-alpha forest, and galaxy clusters. The redshift-space power spectrum and bispectrum of galaxies have some similarity to the summary statistics considered here, with differences including the use of perturbation theory methods for galaxy clustering and the set of nuisance and systematic parameters. Philcox et al. (2021) \citep{fewermocks} and references therein have developed DR approaches for inference with power spectra and bispectra.

Additionally, cross-correlations between the galaxy distribution and CMB secondary anisotropies with galaxy weak lensing can also benefit from our findings. Our results include the cross-correlation with CMB lensing in our analysis. Tomographic cross-correlations with maps of the SZ and ISW effects are also of interest (see forecasts for ``10$\times$2'' and ``12$\times$2'' point analyses of the joint galaxy and CMB observables  \citep{10by2,12by2}). non-Gaussian statistics of these observables may also be considered in the future.

Parameter inference becomes increasingly challenging as one considers longer and more complex data vectors. Regardless of the dataset and analysis framework, CCA and e-MOPED are simple and computationally inexpensive ways to identify the most important parts of the data vector in order to lower computational cost or to ensure robustness in the analysis. We expect they will help enhance the constraints on cosmological parameters with the weak lensing datasets from Stage 4 surveys: Vera C. Rubin Observatory Legacy Survey of Space and Time~\citep{ivezic2019lsst}, Euclid~\citep{laureijs2011euclid}, and the Nancy Grace Roman Space Telescope~\citep{spergel2015wide}.
We leave it for future work to consider the practical applications and generalizations of the techniques presented here.

\vspace{-5pt}

\acknowledgments
We thank Alan Heavens, Mike Jarvis, Helen Qu, Gary Bernstein, Masahiro Takada, Kunhao Zhong, Supranta Boruah, Marco Raveri, Lucas Makinen, Aizhan Akhmetzhanova, Niall Jeffrey for helpful discussions. This work was supported by the OpenUniverse effort, which is funded by NASA under JPL Contract Task 70-711320, ‘Maximizing Science Exploitation of Simulated Cosmological Survey Data Across Surveys’. BJ, MG and MP are partially supported by NASA funding for the Roman and Euclid space telescopes as well.

\appendix




\newpage

\section{PCA-f Constraints}\label{app:pca-f}

\begin{figure}[t!]
    \centering
    \includegraphics[width=0.4\textwidth]{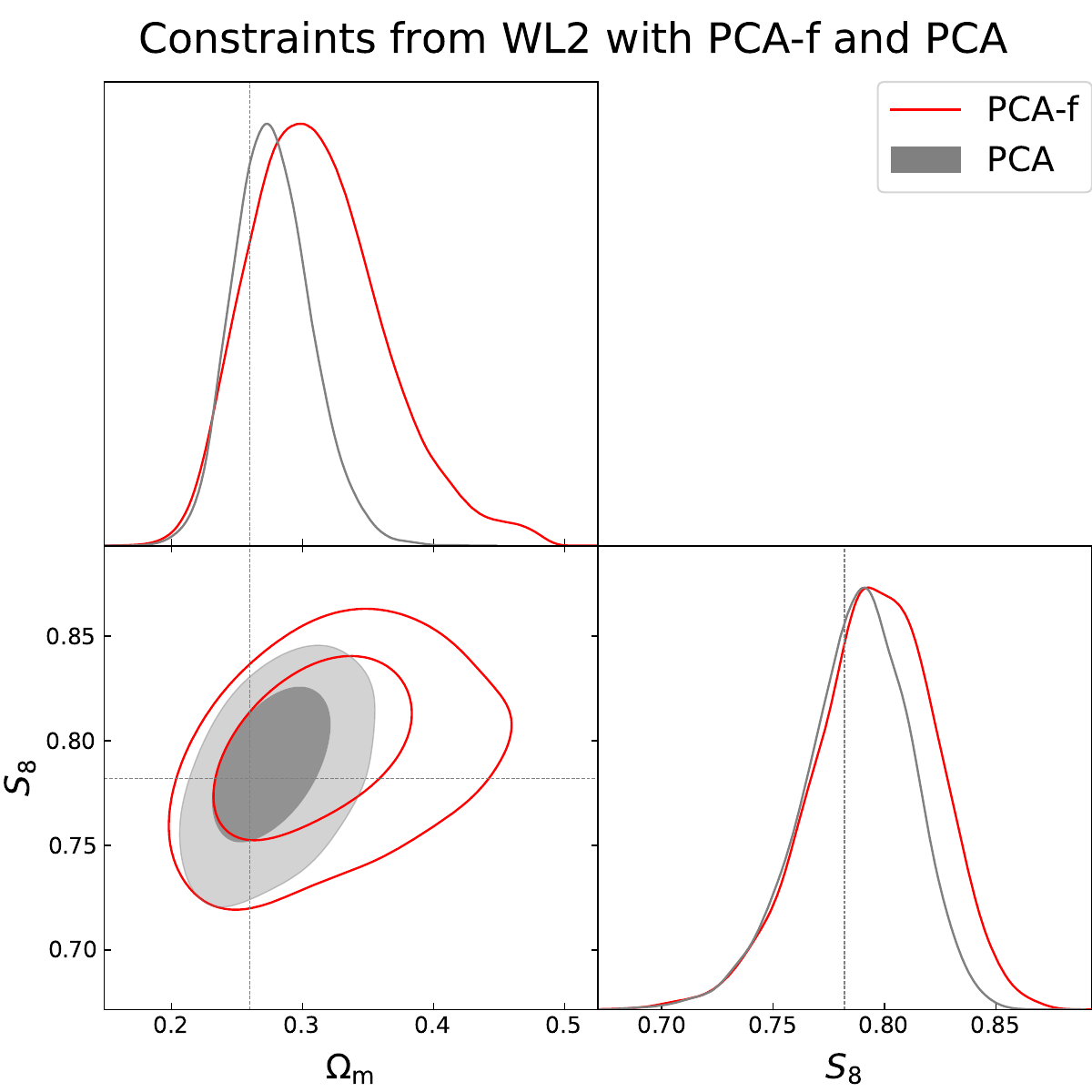}
    \vspace{-10pt}
    \caption{Constraints obtained on $\Omega_m$ and $S_8$ from \wls, with PCA-f (red) and PCA (gray) compression. Though the former is often used, the latter yields much tighter constraints on cosmological parameters. This is because PCA-f compresses the data vector along dominant noise modes, whereas PCA compression can capture variance due to changes in parameters.}
    \label{fig:pcaf}
\end{figure}

As seen in Figure~\ref{fig:pcaf}, the contours produced by PCA-f compression are much broader than those produced with PCA compression. \wls\ is a very high SNR statistic, so a even naive approach like PCA should work well with SBI. However, PCA-f components pick out directions in data space that are uninformative about cosmological parameters. 

\section{Toy Model}\label{app:toy}

\begin{figure}[t!]
    \centering
    \includegraphics[width=0.48\textwidth]{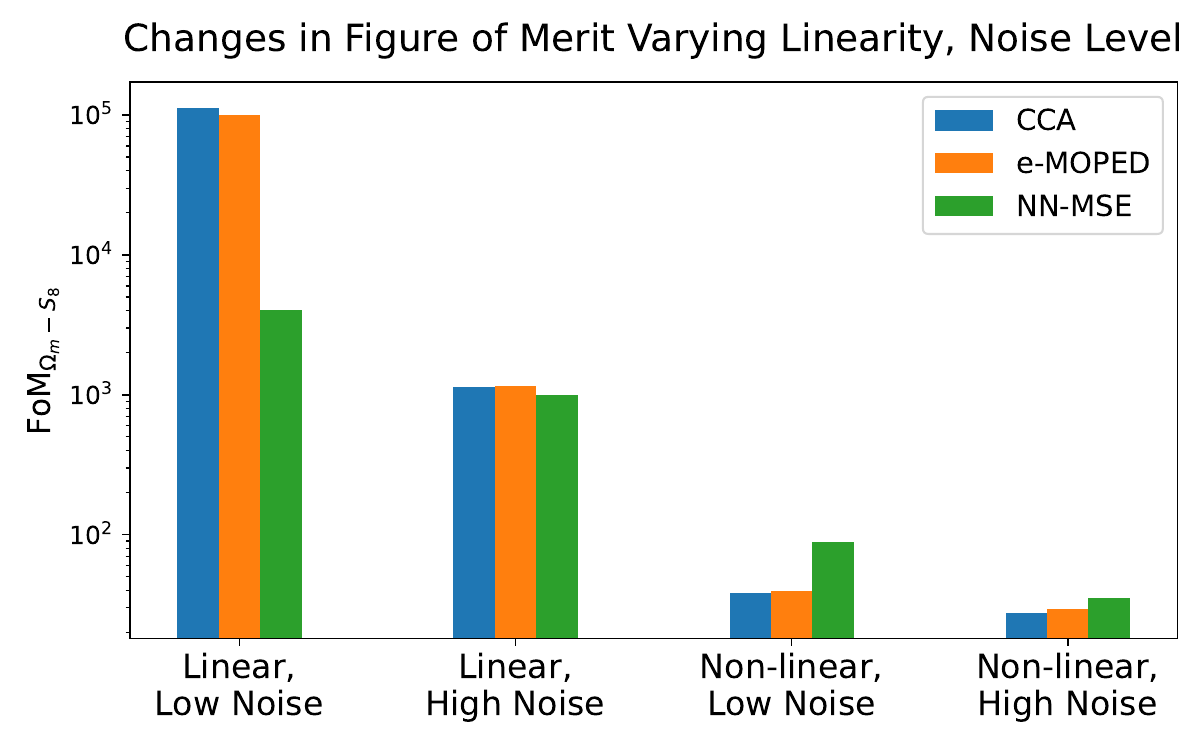}
    \vspace{-10pt}
    \caption{{Results from a toy model of posterior estimation in varying linearity and noise level regimes using CCA, e-MOPED, and NN-MSE. The figures reflect averages over 20 realizations of NDEs. When noise levels are high, the 3 methods perform equivalently well. Otherwise, linear methods are better than NN-MSE when the response to the parameters is linear. } }
    \label{fig:toy}
\end{figure}

{We define a toy model of a posterior estimation problem to identify the regimes in which CCA and e-MOPED might out-perform NN-MSE as we find with weak lensing data vectors.}

{First, we define a 6 dimensional parameter space. The first two parameters will be the parameters of interest, and the rest nuisances. We uniformly sample the $(0,1)^6$ hypercube to populate our simulation prior with $N_{\rm sim}= 20000$ points. Half of the points were used for the compression step, the other half was used for the SBI after compression. This mimics realistic $N_{\rm sim}$ and parameter space for weak lensing SBI scenarios. }

{Then we define 100 dimensional linear and non-linear data vectors. For the linear data vector, we define an arbitrary ${100\times 6}$ matrix $A$ where every element is between $-1$ and $1$ such that $Ap$ is the noiseless linear data vector where $p$ is a given parameter vector. For the non-linear data vector, we define an arbitrary ${100\times 6}$ matrix $B$ where every element is between $2$ and $6$. Then, the $i^{\rm th}$ element of the noiseless non-linear data vector $v$ is defined as 
\begin{align}
    v_i = \sum_{j=1}^6 p_j^{B_{i,j}^{\ }}
\end{align}
where $p_j$ is the $j^{\rm th}$ parameter. }

{The noise in our analysis was modeled as additive white noise, meaning it was independent across data vector components and parameters, with a defined standard deviation. For linear data vectors, we used two noise levels: low noise with $\sigma_{\rm low}=0.02$ and high noise with $\sigma_{\rm hi}=0.2$. When working with non-linear data vectors, we chose their $\sigma_{\rm low}$ and $\sigma_{\rm hi}$ values to ensure that the determinant of the Fisher Information Matrix (calculated for the first two parameters) matched the corresponding values from the linear case at each noise level. This approach allowed us to maintain consistent noise conditions when comparing linear and non-linear models. Practically speaking, this is just a constant rescaling of $\sigma_{\rm low}, \sigma_{\rm hi}$ as a function of $A$ and $B$. Note that this makes the noise levels fair at the level of the Cramer-Rao bounds, and does not guarantee comparable FoM.}

{With these data vectors, we use CCA, e-MOEPD, and NN-MSE to perform SBI on the first two parameters given a data vector generated at $p_{\rm fid} = (0.5,0.5,0.5,0.5,0.5,0.5)$. The resulting FoM are shown in~\Cref{fig:toy}. We find that when noise levels are high, the choice of DR is not very important. We also find that when noise levels are low, the choice of DR  impacts FoM. If the response to parameter is non-linear, NN-MSE is clearly better. On the other hand, when the response to parameter is linear, linear methods out-perform NN-MSE. }

{An important note here is that response to parameter is always linear given a small enough simulation suite prior. Even if the true response is non-linear (as is the case with $\sigma_8$ and \texttt{WPH}, small scale \wls, etc.), if the simulation suite prior is small enough the response to parameter is linear to dominant order. }

{This gives some context to our findings that CCA and e-MOPED out-perform NN-MSE. This toy exercise would suggest that even with the non-Gaussian statistics in \all, DES Y3 level weak lensing SBI is in a regime where 
\begin{enumerate}
    \item noise levels are low enough for choice of DR to matter, and
    \item the response to parameter is linear enough within the simulation suite prior that linear methods of DR are better than NN-MSE.
\end{enumerate} 
From this one might speculate that with future surveys with lower noise, the choice of DR will become even more important. Furthermore, the Bayesian nature of cosmology affords future surveys tighter simulation suite priors, making linear methods even more of valid choice over NN-MSE. }

\section{Sensitivity to Scale Cuts}\label{app:scalecut}
\begin{figure}[t!]
    \centering
    \includegraphics[width=0.48\textwidth]{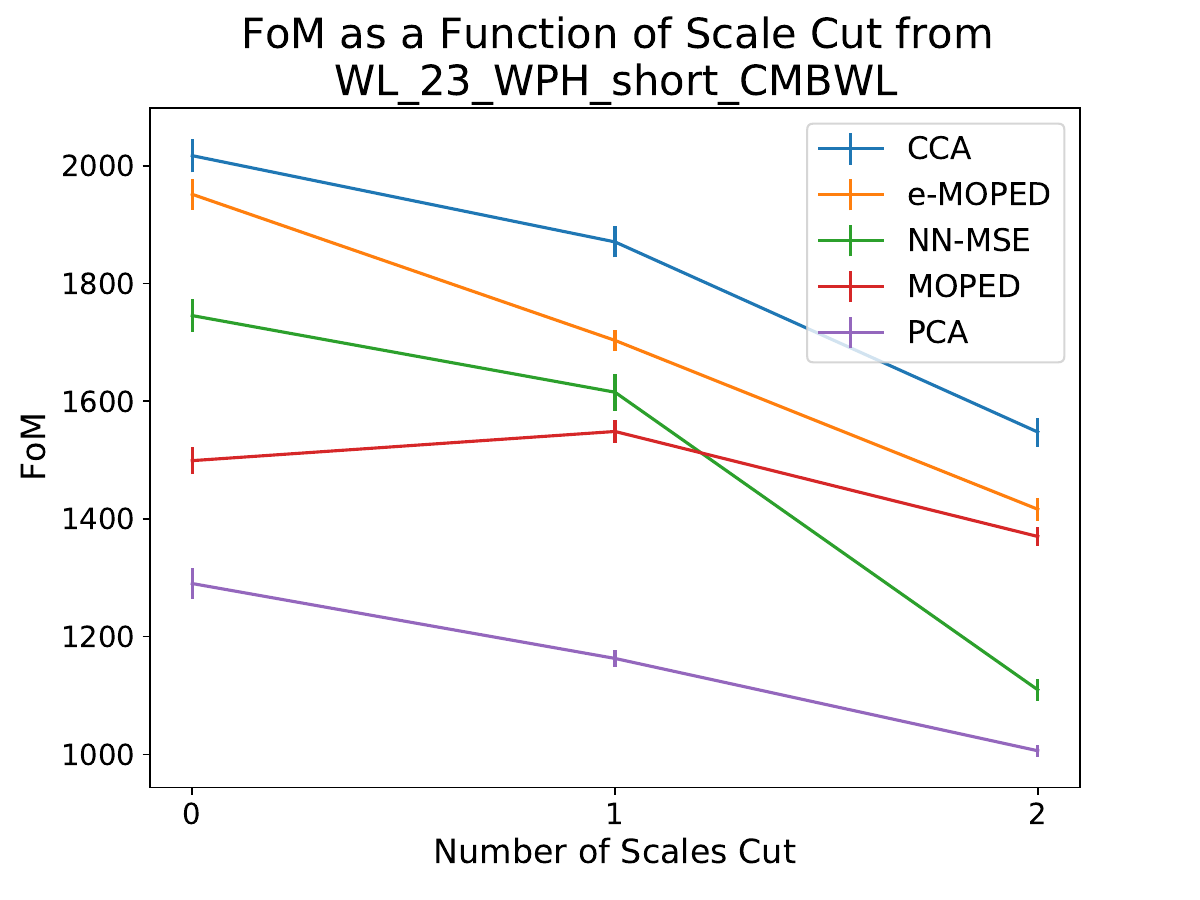}
    \vspace{-20pt}
    \caption{{FOM obtained with \all\ while varying the number of scales cut from the smallest scales. The curves reflect the mean of 20 NDE realizations, and the error bars are the standard error. The conclusion that CCA (blue) and e-MOPED (orange) are the best out of the methods showcased is robust to scale cuts.  NN-MSE (green) and MOPED (red) fail to efficiently capture the information at the smallest scale due to the low SNR. }  }
    \label{fig:scalecut}
\end{figure}

{To probe the sensitivity of our results to scale cuts, we repeated the exercise of computing the FOM in $\Omega_m - S_8$ plane with \all\ using CCA, e-MOPED, NN-MSE, MOPED, and PCA while varying the number of smallest scales cut. We stop at cutting the two smallest scales, because further cuts are too restrictive for the \texttt{WPH} portion of the data vector.} 

{We find that CCA and e-MOPED (in that order) outperform all other methods across different scale cuts. The bend in the NN-MSE curve reflects the fact that the smallest scale is too noisy for neural networks to efficiently extract information. FOM from MOPED appears to increase with the first scale cut, because, as mentioned elsewhere, the performance of MOPED hinges on the convergence of numerical derivatives. Numerical derivatives of non-Gaussian statistics at the noisiest scales is extremely noisy and renders MOPED ineffective overall. This highlights the point that NN-MSE and MOPED are ill suited to compressing low SNR statistics. }

\section{Correlation of $c_i$ and $p_i$}\label{app:cov}

\begin{figure*}[t!]
    \centering
    \includegraphics[width=0.49\textwidth]{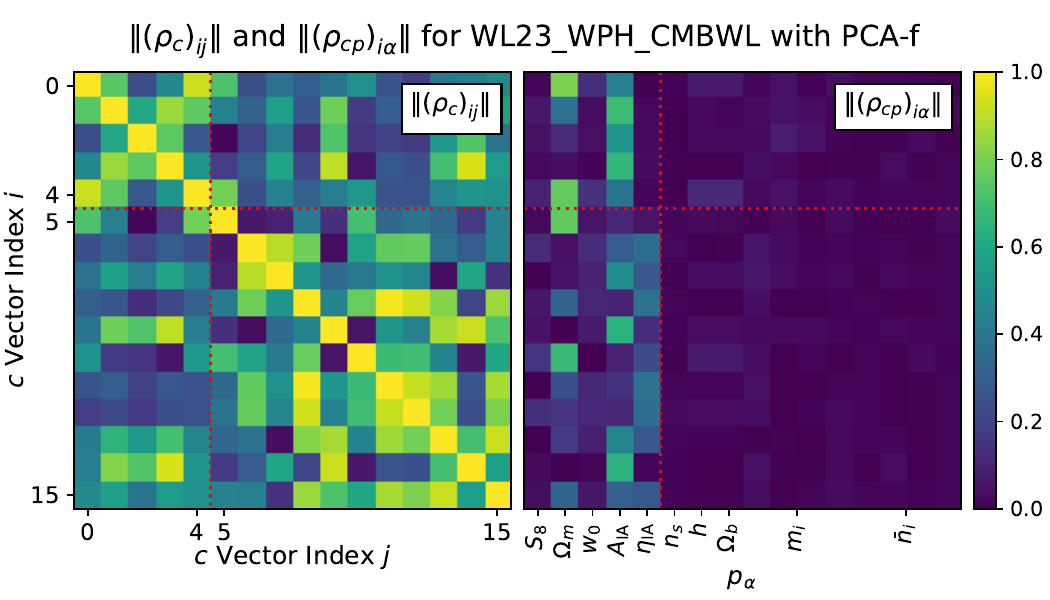}
    \includegraphics[width=0.49\textwidth]{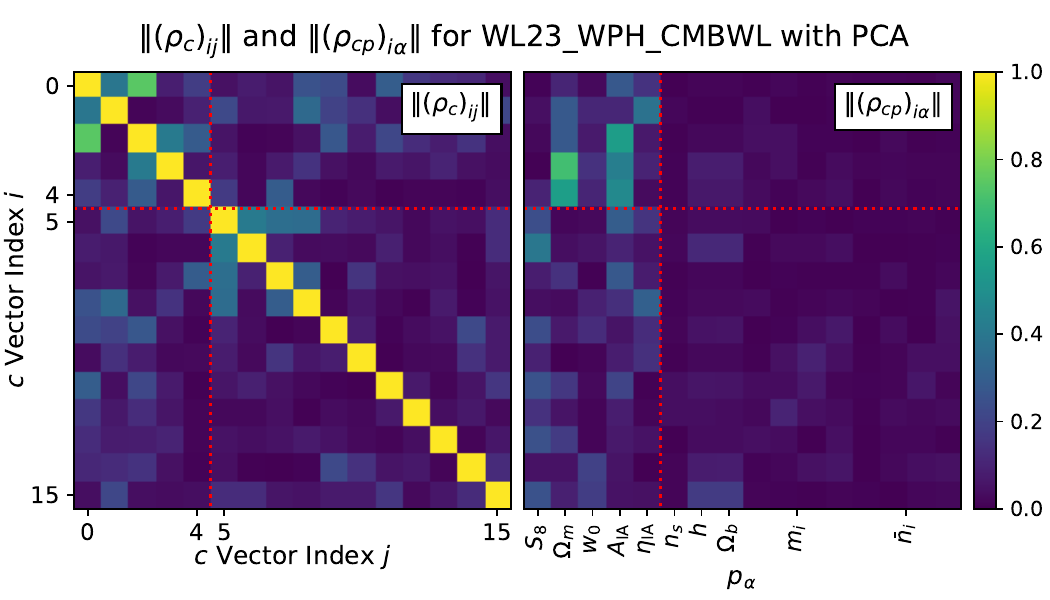}
    \includegraphics[width=0.49\textwidth]{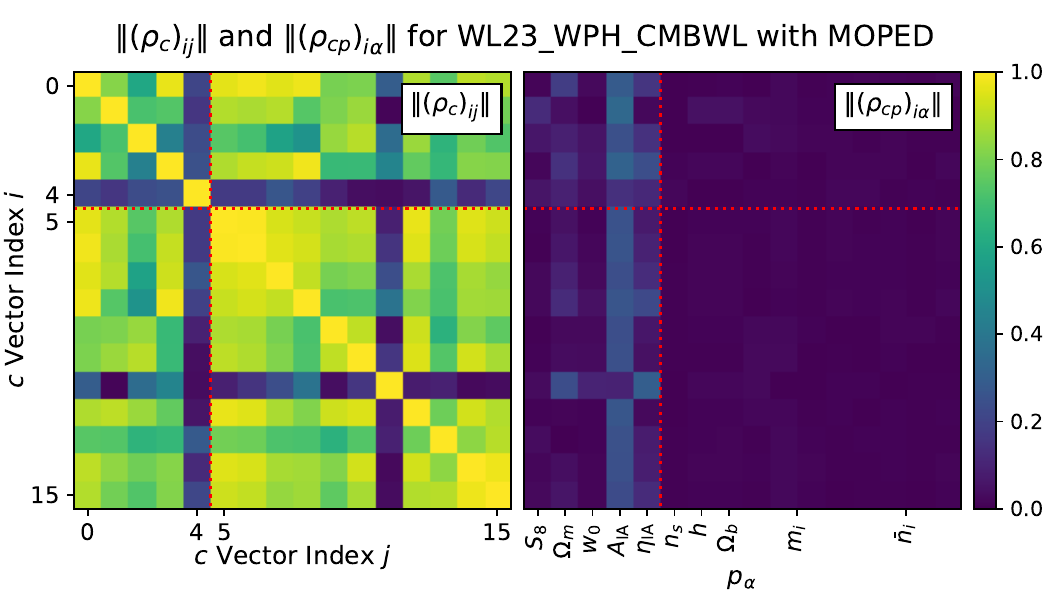}
    \includegraphics[width=0.49\textwidth]{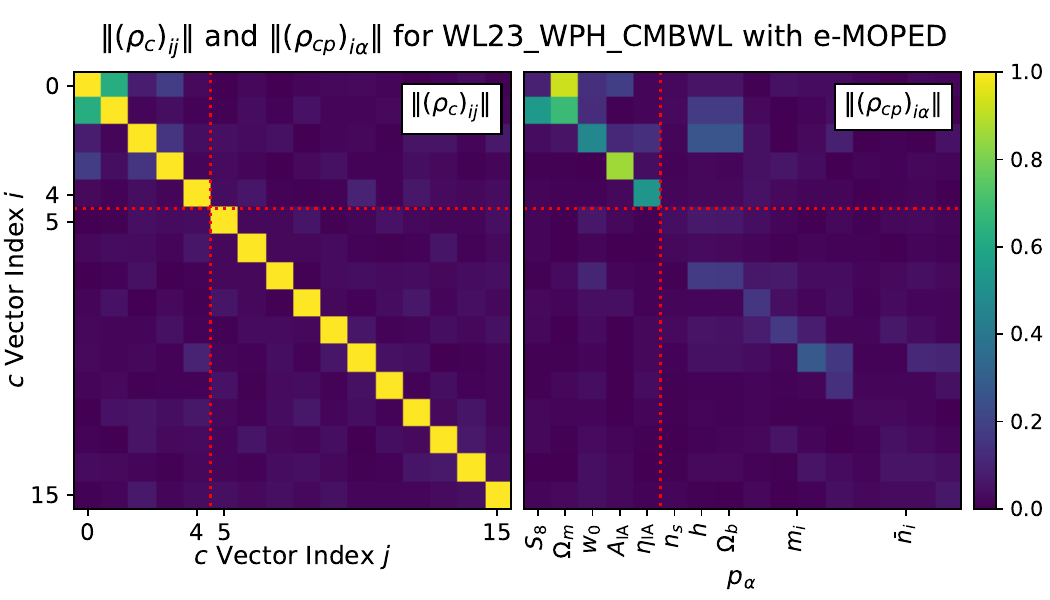}
    \includegraphics[width=0.49\textwidth]{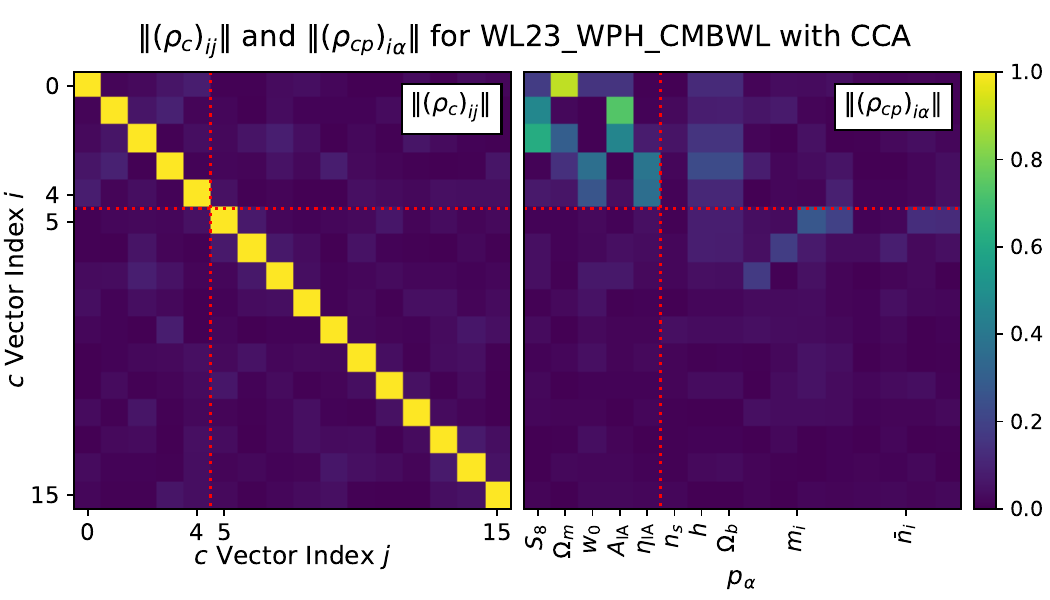}
    \includegraphics[width=0.49\textwidth]{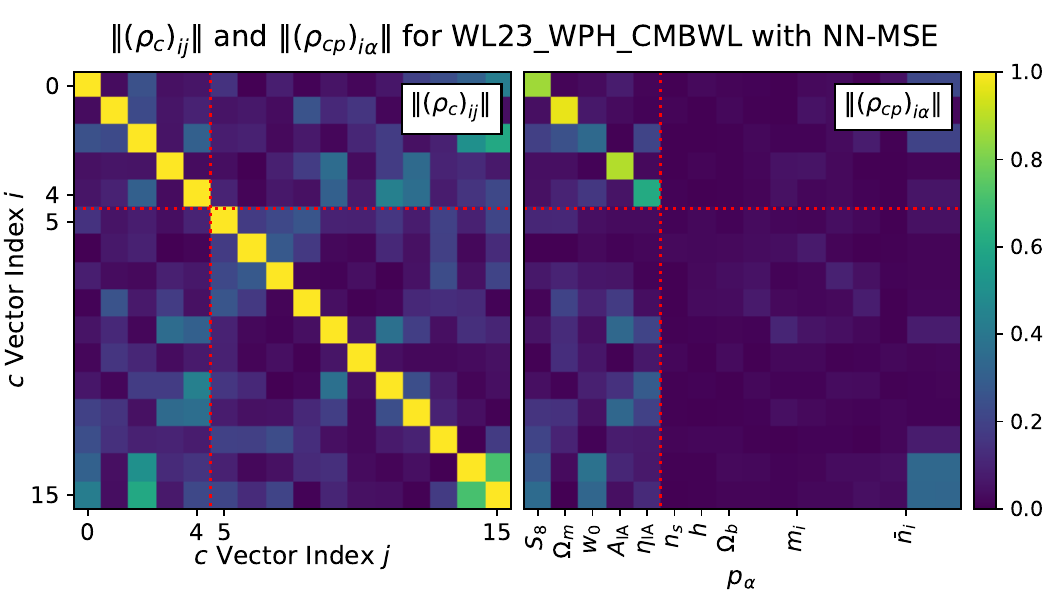}
    \vspace{-10pt}
    \caption{$\rho_{c}$ is the auto-correlation matrix for elements of $c_i$ and $\rho_{cp}$ is the cross-correlation matrix between elements of $c_i$ and $p_i$. The matrices are displayed as the absolute values of their elements ($\abs{(\rho_c)^{\ }_{ij}}$ on the left and $\abs{(\rho_{cp})^{\ }_{ij}}$ on the right), as it is the strength of the statistical relationship that is relevant. The full \all\ data vector was compressed with PCA-f (top left), PCA (top right), MOPED (center left), e-MOPED (center right), CCA (bottom left), and NN-MSE (bottom right) to produce the figures. The red lines indicate the most important $5\times5$ block corresponding to $\rho_c$ and $\rho_{cp}$ of the first 5 elements of $c_i$ and ($S_8,\Omega_m, w_0, A_{\rm IA}, \eta_{\rm IA}$). A desirable compression scheme will have two things. First: a roughly diagonal left panel, indicative of the statistical independence of elements of $c_i$. Second: large values in the first two columns of the second panel, indicating informativeness about $S_8$ and $\Omega_m$.  }
    \label{fig:cov}
\end{figure*}

One way to gauge the performance of a compression method before computing the FoM is by examining the correlation matrices of $c_i$ and $p_i$.~\Cref{fig:cov} shows $\abs{(\rho_c)^{\ }_{ij}}$ and $\abs{(\rho_{cp})^{\ }_{ij}}$ for \all\ using PCA, PCA-f, MOPED, e-MOPED, CCA, and NN-MSE, where $\rho_c$ is the auto-correlation matrix of $c_i$ and $\rho_{cp}$ is the cross-correlation matrix of $c_i$ and $p_i$. We show the absolute values of the elements of these correlation matrices. A good compression method should result in data vectors with uncorrelated elements so that each element adds new information, and data vectors highly correlated with the parameters of interest $S_8$ and $\Omega_m$. The first translates to a diagonal $\rho_c$ and the second translates to the first two columns of $\rho_{cp}$ being non-zero.

Though these correlation matrices do not tell the whole story, it is easy to tell from them why certain methods perform better than others. PCA results in a much more diagonal $\rho_c$ than PCA-f. The first row of $\rho_{cp}$ from PCA-f is nearly zero. MOPED results in a very non-diagonal $\rho_c$ and the first two columns of $\rho_{cp}$ from e-MOPED are further from zero. From this we would be able to guess that PCA and e-MOPED outperform PCA-f and MOPED respectively. 

e-MOPED, PCA, CCA, and NN-MSE all look quite good in the two critera mentioned above. NN-MSE results in a slightly less diagonal $\rho_c$ but discerning the performance of these three would require performing the full SBI. 

Note that, mathematically speaking, $\rho_c$ from PCA should be diagonal. However, due to differences in the sample of data vectors used to generate the Principal Components and the  sample used to generate the correlation matrix, the off diagonals are non-zero.

\bibliographystyle{apsrev4-1}
\bibliography{biblio}

\end{document}